# Evolution of the hydrogen-bonded network in methanol-water mixtures upon cooling

Ildikó Pethes[1,a], László Pusztai[a,b], László Temleitner[a]

[a]Wigner Research Centre for Physics, Konkoly Thege út 29-33., H-1121 Budapest, Hungary
[b]International Research Organization for Advanced Science and Technology (IROAST), Kumamoto University, 2-39-1 Kurokami, Chuo-ku, Kumamoto 860-8555, Japan

**Abstract**

The hydrogen-bonded structure of methanol – water mixtures is investigated over the entire alcohol concentration range (from $x_{\text{Methanol}} = 0.1$ to $1.0$) at several temperatures, from 300 K down to the freezing point of the given mixture. Classical molecular dynamics simulations have been carried out, using the all-atom OPLS-AA force field for methanol and the TIP4P/2005 model for water molecules. Simulation trajectories ('particle configurations') obtained have been analyzed, in order to characterize the hydrogen-bonded network in the mixtures. The temperature and concentration dependence of the average hydrogen bond (H-bond) numbers between different types of molecules, the donor/acceptor roles of water and methanol molecules, and hydrogen bond number distributions have been revealed. The topology of the total system, as well as that of the water and methanol subsystems, has been investigated by calculating the cluster size distributions, the number of primitive rings, and ring size and ring type distributions. It has been found that upon cooling, the average number of H-bonded water molecules increases at every concentration and temperature investigated. As far as the connectivity of the hydrogen-bonded network is concerned, the percolation threshold has been shown to be above $x_M = 0.9$ already at room temperature.

*Keywords: molecular dynamics simulation, methanol-water mixtures, temperature dependence, hydrogen bond network*

---

[1]Corresponding author: e-mail: pethes.ildiko@wigner.hu

# 1. Introduction

It is our pleasure to greet Professor Myroslav Holovko on the occasion of his 80[th] birthday. Professor Holovko has made a number of important contributions to the theory of aqueous solutions (see, e. g. [1, 2]). It is this line of his research that we wish to connect to by the present contribution.

Hydrogen bonds (H-bonds) play an outstanding role in many physical, chemical, and biological processes and determine the structure and properties of many molecular liquids, solids, and biomolecules: among others, those of water, alcohols, proteins, sugars, or DNA. Molecules can be connected via H-bonds and can form supramolecular structures called 'H-bonded network'-s.

H-bonding is considered to be responsible for several anomalous properties of water, e. g. its high boiling and melting points, viscosity, or the crystal structure of ice (see e. g. [3 – 6], and references in there). Theoretically, water molecules can establish four H-bonds: two as H acceptors through the lone electron pairs of their O atoms, and two as H-donors by their H atoms. Fourfold coordinated water molecules are tetrahedrally arranged and form a percolating, three-dimensional H-bonded network. This network consists of cycles (chains that are closed into themselves), among which the 5 and 6-membered rings are the most frequent [7].

Alcohol molecules possess hydrophilic (OH groups) and hydrophobic (alkyl groups) parts. The OH group of the alcohols takes part in H-bonds (though there are suggestions about the presence of a weak H-bond with the participation of the $CH_3$ group, see e. g. [8 – 11]). Monohydroxy alcohols have one OH group, with two H acceptors, but only one H donor option; therefore, they can form at most two H-bonds on average in pure alcohols. Alcohol molecules form linear chains or small cyclic aggregates [7, 8, 12 – 15].

Methanol is the prototype of alcohols, and also a close analogue of water, in which one proton is replaced by a methyl group. However, the structure of pure methanol is significantly different from the structure of water, caused by the influence of hydrophobic methyl groups and the smaller number of possible H-bonds [7, 8, 15, 16].

In alcohol – water mixtures, H atoms of water become available also to alcohol molecules and thus, alcohol molecules can participate in more than 2 (up to 3) H-bonds. The average number of H-bonds depends on concentration, temperature, and pressure. In these systems, many new aspects of the developing H-bonded network appear, such as the binding preferences between individual molecular types, the percolation of the network and its subsystems, or the size and type distribution of the rings formed: quite a few publications deal with this topic (for some recent works, see, e. g. [17 – 34]). In general, it has been observed that at low alcohol concentrations, the three-dimensional H-bonded network of water dominates, in which alcohol molecules are independently hydrated. As the alcohol concentration increases, chain-like structures formed mostly by alcohol molecules become more abundant. At high alcohol concentrations and at room temperature, the H-bonded network ceases to percolate: in ethanol – water mixtures this occurs above $x_{Ethanol}$ = 0.9 [22], and in 2-propanol – water mixtures above $x_{2\text{-Propanol}}$ = 0.6 alcohol concentrations [24]. At lower temperatures, the average number of H-bonds increases and the connectivity of the network becomes stronger, so percolation can be observed up to higher alcohol concentrations [22, 24].

The temperature dependence of the H-bonded structure in pure methanol and methanol – water mixtures has been investigated in some publications earlier, by different simulation (and simulation plus experimental) techniques. As it is obvious from the short overview below, there are still a number of unsettled issues. It was reported that the average size of clusters in pure methanol increases with decreasing temperature [35, 36]. In contrast, the average number of H-bonds was found to be independent of temperature, using a purely energetic definition of H-bonds [37]. The authors found that the probability of 2-fold H-bonded methanol molecules increases with decreasing temperature, causing the reduction of the number of branches. Using a purely geometric definition of H-bonds, higher average H-bond number was reported at lower temperatures [36]. The increase of the average H-bond number was also observed in our previous study, using either purely geometric or a combined (energetic plus geometric) definition of H-bond [28].

NMR investigations showed concentration and temperature dependence of the favored types of H-bonds in methanol – water mixtures [38 – 40]. At higher temperatures, H-bonds between methanol and water molecules are preferred. As the temperature decreases, H-bonds between like molecules (methanol – methanol and water – water) are more frequent. The crossover temperature decreases with increasing alcohol concentration.

Larger water clusters were reported in methanol – water mixtures upon cooling and by applying higher pressures, as obtained from molecular dynamics (MD) simulations [41]. From their X-ray diffraction measurements, Takamuku and coworkers suggested that in water-rich mixtures, the composition of the dominant clusters is not changing on cooling [42]. An increase of the average number of H-bonds and the probability of 5 and 6-membered rings in water-rich mixtures was found by MD simulations [34].

Several small (up to 11 molecules) methanol – water clusters were investigated by density functional theory calculations [43] in the temperature range of 273 to 323 K. These computations showed that mixed clusters dominate in the $0.1 \leq x_M \leq 0.8$ concentration region; the maximum population of mixed clusters was found at $x_M$ = 0.365 (at $T$ = 273 K). Above this concentration, the self-association of methanol clusters becomes more preferred as the methanol concentration increases. Note, however, that cluster calculations may not be fully relevant with respect to the bulk phase.

Our group has previously investigated the H-bonded network of several alcohol – water systems upon cooling (methanol – water [28, 34], ethanol – water [22, 29, 30], 1-propanol – water [23], 2-propanol – water [24, 31]). Concerning methanol – water mixtures, the network was analyzed at low methanol concentrations only [34]. Galicia et al. [44] have complemented their room temperature X-ray diffraction measurements with molecular dynamics computer simulations over the entire composition range. In Ref. [28] X-ray and neutron diffraction data were reported for the whole concentration range of methanol – water mixtures, from room temperature to the freezing point. In that paper, it was observed that classical MD simulations using the OPLS/AA (for methanol) [45] and TIP4P/2005 (for water) [46] potential models result in particle configurations that satisfactorily reproduce the measured structure factors. It was found that the average number of H-bonds increases upon cooling: this tendency was most significant concerning bonds between unlike (water and methanol) molecules. At the same time, the number of H-bonds slightly decreases between water molecules above 70 mol% methanol content, as well as between methanol molecules in the 30 – 60 mol% methanol concentration range.

The present study is a continuation and extension of our previous work [28], in which the H-bonded environment of molecules of methanol – water mixtures is studied over the entire alcohol concentration range (from $x_M = 0.1$ to 1.0), at several temperatures from 300 K to the freezing point of the mixture. Apart from refining some of the previous findings, here we focus more on the properties of the entire hydrogen bonded network.

A technical note concerning the Supplementary Material (SM) that accompanies the main text: as our aim was to provide a picture on the temperature-composition dependence of H-bonding properties in methanol-water mixtures, a rather extensive set of figures have been generated. The main text contains and discusses each figure that is connected to the itemized conclusions. Figures in the SM are not discussed in the main text, in order to facilitate an uninterrupted reading: instead, the existence of the supplementary figures is just briefly indicated wherever it is relevant.

**2. Simulation method**

Classical molecular dynamics (MD) simulations have performed by the GROMACS software package (version 2018.2) [47]. Water molecules were represented by the TIP4P/2005 [46] model, and the all-atom OPLS-AA [45] model was used for methanol: this combination of interatomic potential functions has been shown to work satisfactorily not only in structural studies [28], but also, when considering the apparent molar volume in the same kind of mixtures [48]. Cubic simulation boxes contained 2000 molecules. The molecules were initially randomly placed into the simulation box. Starting box sizes were chosen according to the room temperature densities from Ref. [49]. In the first step, energy minimization was performed using the steepest-descent method; after that, the equations of motion were integrated via the leapfrog algorithm. The time step was 2 fs. A heat treatment was applied to avoid the aggregation of molecules, during which a 2.2 ns long simulation run was performed at 320 K. In the next step, the simulation box was cooled down to the investigated temperature ($T_{inv}$), using the 'simulated annealing' option of the GROMACS software, with a 0.2 K / ns cooling rate. At the investigated temperature, 2 equilibration steps were carried out: a 10 ns long NpT run was followed by a 2 ns long NVT (constant volume and temperature) run. The production run was 20 ns long; the trajectories were saved at every 200 ps. (In these steps, the last saved simulation box of the previous step was used as starting configuration.) Simulation stages are collected in Table 1, together with the applied thermostats and barostats. For a detailed description of the force fields and other parameters of the simulations, see our previous work [28].

The concentration – temperature points considered are collected in Table 2, together with the simulation box sizes applied (in the production runs, at the given temperature).

101 configurations, 200 ps apart, were collected for the purpose of examining H-bond related characteristics. H-bonds were localized by a geometric definition: two molecules were identified as H-bonded if the intermolecular distance between an oxygen and a hydrogen atom was less than 2.5 Å, and the corresponding O...O-H angle was smaller than 30 degrees. Calculations concerning H-bonds and H-bonded networks were performed by an in-house software, based on the HBTOPOLOGY code [16, 55]. (Note that all calculations have been repeated using a combined geometric – energetic definition of H-bonds: the O...H distance needed to be less than 2.5 Å, and the interaction energy had to be smaller than -12 kJ/mol for an H-bond. It has been found that although the actual values of H-bond numbers slightly differ between the two different H-bond definitions, the tendencies and overall results are identical.)

## 3. Results and discussion

*Number of H-bonds*

The average number of H-bonds ($N_{Hb}$) and the average number of H-bonded molecules around water molecules ($N_W$) increase monotonously with decreasing temperature, see Fig. 1, where the numbers are normalized by the 300 K values. The increase is around 13% at the lowest temperature investigated. (It is similar to what was previously found in ethanol – water, 1-propanol – water, and 2-propanol – water mixtures [22 – 24].) It should be noted here that the increase of $N_{Hb}$ becomes slower at lower temperatures, where it approaches the theoretical maximum average values ($2*x_M+4*(1-x_M)$). The average H-bond number of methanol molecules ($N_M$) increases with decreasing temperature at low ($x_M \leq 0.4$) and high ($x_M \geq 0.9$) methanol concentrations, and also between these values down to about 200 K. However, at the lowest temperatures, for methanol concentrations $0.5 \leq x_M \leq 0.8$, the increase stops, and even a slight decrease can be observed. A maximum can be observed in the number of H-bonds in this medium concentration range. This behavior was not previously reported in alcohol – water mixtures: in ethanol – water [22] and 1-propanol – water [23] mixtures, the average number of H-bonds around the alcohol molecules also increases as the temperature decreases.

Concerning the different molecular pairs, the number of H-bonds between unlike molecules increases with decreasing temperature over essentially the entire temperature and concentration range investigated. The increment is around 20% at the lowest temperatures tested (see also Fig. 1). The highest increase in terms of the number of H-bonds between water molecules ($N_{WW}$) is observed at low methanol concentrations ($x_M \leq 0.3$). As the ratio of water molecules decreases in the mixtures, $N_{WW}$ increases more slowly and even decreases at high methanol concentrations and low temperatures ($0.6 \leq x_M \leq 0.8$, $T < 233$ K; note that statistics for the $x_M = 0.9$ mixture are so bad that the validity of the peculiar behavior of the corresponding curve may be questionable). The number of H-bonds between methanol molecules ($N_{MM}$) mainly decreases with decreasing temperatures at low methanol concentrations ($x_M \leq 0.6$), showing a maximum value (around 200-230 K) in the $0.7 \leq x_M \leq 0.8$ concentration range, and increases monotonously at high ($x_M > 0.8$) methanol concentrations. Interestingly, the largest increase in terms of the number of methanol-methanol H-bonds ($N_{MM}$, around 12.7%, in pure methanol) is more than the maximum increase of the number of H-bonds between water molecules ($N_{WW}$, 8.2% at $x_M = 0.4$).

A similar behavior of $N_{WW}$ has been observed in other alcohol – water mixtures: $N_{WW}$ increases with decreasing temperature in water-rich mixtures, but it is independent of temperature, or even slightly decreases in alcohol-rich mixtures [22, 23]. An increase in terms of the number of alcohol – alcohol H-bonds can also be observed in other alcohol-rich mixtures [22, 23]. Temperature independent alcohol – alcohol H-bond numbers were reported in 2-propanol – water mixtures for the entire concentration range [24] and in water-rich alcohol – water mixtures in ethanol – water [22] and 1-propanol – water mixtures [23]. However, no decrease in terms of the number of alcohol – alcohol H-bonds could be observed previously.

This comparison shows that as temperature decreases, the increasing average number of H-bonds ($N_{Hb}$, between any molecules) is mainly the result of the increasing number of H-bonds between unlike molecules.

We note, in passing, that the un-normalized values have also been reported in our previous work [28], and that they agree qualitatively with those reported in the earlier work of Galicia et al. [44]. (For more details, please see Figures S1 – S4 in the SM.)

*H-bond number distributions*

The distribution of the number of H-bonds at different temperatures and methanol concentrations is shown in Fig. 2. Concerning the entire system, at low methanol concentrations, 4-fold H-bonded (water) molecules dominate, while at high methanol concentrations, (mostly methanol) molecules with 2 H-bonds are the most frequent. Around $x_M = 0.5$, molecules with 2 and 4 H-bonds are similarly popular as the molecules with 3 H-bonds. As the temperature decreases, the fraction of molecules with 0 or 1 H-bonds decreases.

The number of water molecules with 4 H-bonds increases similarly for all water concentrations with decreasing temperature, along with the deterioration of the number of water molecules with 2 H-bonds. The occurrence of 3-fold H-bonded water molecules decreases, in general, with decreasing temperature. However, it shows a slight maximum in the $x_M = 0.9$ mixture at $T = 268$ K.

Methanol molecules can establish their $3^{rd}$ H-bonds in the presence of available water molecules. Thus the probability of methanol molecules with 3 H-bonds is higher at low methanol concentrations, and the number of these molecules increases with decreasing temperature. At high methanol concentrations, cooling causes an increase in the number of the 2-fold H-bonded methanol molecules. More details can be found in the SM (Figs. S5 – S7).

*Donor – acceptor states*

Water and methanol molecules can participate in H-bonds as proton donors (D) and acceptors (A); proton donor molecules are defined as molecules that are connected by their hydrogen atoms. Molecules are said to be in the '$N_D$ D – $N_A$ A' state if they participate in $N_D$ H-bonds as donors and $N_A$ H-bonds as acceptors. The relative frequencies of the most popular donor – acceptor states of water and methanol molecules at 300 K and at the lowest investigated temperature are shown in Figs. 3 and 4. (A more detailed temperature and concentration dependence can be found in the SM, Figs. S8 and S9.)

The most frequent D – A states of water molecules are the 2D – 2A and 2D – 1A states: the former is the most frequent one at low alcohol concentrations, while the latter is at high methanol concentrations. There are considerable amounts of molecules in the 1D – 1A, 1D – 2A, and 2D – 0A states at room temperature, which states are less frequently taken at low temperatures. As temperature decreases, the popularity of the 2D – 1A state also diminishes in favor of the 2D – 2A state.

The most popular methanol D – A state at room temperature is the 1D – 1A one, followed by the 1D – 2A, 0D – 1A, and 1D – 0A states. The 1D – 2A state is more frequent at low methanol concentrations, while the number of molecules with 1 H-bond (0D – 1A and 1D – 0A states) is higher at high methanol concentrations. Similarly to water, D – A states with a higher number of H-bonds become more popular as temperature decreases, and the importance of 1D – 0A, 0D – 1A, 0D – 0A states decreases. The temperature dependence of the 1D – 2A state shows a maximum for $x_M \geq 0.3$.

It could also be observed in other alcohol – water mixtures that donor – acceptor states with more H-bonds become more popular as temperature decreases. The 2D – 2A and the 2D – 1A states of water and the 1D – 1A state of alcohol molecules were reported as the most preferred states [22 – 24].

The concentration and temperature dependence of the average number of H-bonds of water and methanol molecules as donors and as acceptors ($N_W^D$, $N_W^A$, $N_M^D$, and $N_M^A$) is presented in Fig. 5. As temperature decreases, the number of donor H-bonds of both methanol and water molecules increase and tend to the theoretical maximum value (for methanol 1, for water 2). Around 200 K, their values are close to the maximum value (within 2% for methanol and 0.5% for water). As the number of donor molecules saturates, the increase in the number of H-bonds becomes slow, and less and less new H-bonds are formed.

The behavior of the number of acceptor molecules is different for methanol and water. In the case of water, the number of acceptor molecules increases nearly linearly with decreasing temperature at every methanol concentration and over the entire temperature range investigated (without significant changes). As temperature decreases from room temperature, $N_M^A$ starts to increase. However, on further decrease of temperature, $N_M^A$ reaches a maximum value and starts to decrease. This can be clearly seen for methanol concentrations in the $0.3 \leq x_M \leq 0.8$ range. For low ($x_M < 0.3$) and very high ($x_M \geq 0.9$) methanol concentrations, such maximum cannot be seen.

As H-bonds between like molecules affect the number of acceptor and donor bonds equally, the difference in terms of the behavior of methanol and water acceptors originates from the methanol – water H-bonds. At this point, it becomes clear that the two types of water – methanol H-bonds must be distinguished. It is also supported by the energy distribution of the different H-bonded pairs (see e. g. Fig. 6 for $x_M = 0.6$ at $T = 300$ K and 178 K). The water(D) – methanol(A) (WD-MA) pairs have the lowest (bonding) energy, followed by the water – water, and methanol – methanol pairs, and the methanol(D) – water(A) (MD-WA) pairs have the highest energy. We note that such behavior of the energy distribution of alcohol – water pairs was also reported in 2-propanol – water mixtures [24].

The temperature dependence of the average number of H-bonded A and D methanol molecules around water (per water molecule, $N_{WD-MA}$, and $N_{WA-MD}$), normalized to the 300 K values, is shown in Fig. 7.

Both $N_{WD-MA}$, and $N_{WA-MD}$ increase with decreasing temperature, but the growth is significantly greater for pairs where methanol molecules act as donors (40% vs. 10%). This can be taken as a manifestation of the fact that 'water D – methanol A' type bonds are very stable already at room temperature (cf. Fig. 6).

Comparing all this with the tendencies of $N_{MM}$, $N_M$, and the number of methanol molecules in the 1D – 2A state, the following scenario may be drawn:

At room temperature, all the 4 possible types of H-bonds are formed with different probabilities: methanol(D) – methanol(A), methanol(D) – water(A), water(D) – methanol(A), water(D) – water(A). The H atoms of water molecules (water donors) are in H-bonds most frequently (90 – 95% of them), while only 82 – 86% of the H atoms of methanol molecules participate in H-bonds (the higher numbers belong to the higher methanol concentrations). The same numbers for the acceptor possibilities of the molecules are 60 – 90% for water and 45 – 70% for methanol (higher values are connected to lower methanol concentrations). The H-bonded pairs with deeper potential energy are more frequent: these are water(D) – methanol(A) and water – water pairs (cf. Fig. 6). As temperature decreases, the possible donor states are occupied with a higher probability, and more pairs are trapped in the potential wells. Below 250 – 200 K, the number of newly formed H-bonds decreases: particularly the number of new water D states becomes very low. Meanwhile, the number of water A states increases at a similar rate, which means that the newly formed bonds below this temperature are mostly the type methanol(D) – water(A). Moreover, below a certain temperature (that depends on the methanol concentration), the number of methanol – methanol H-bonds ($N_{MM}$) decreases, albeit the number of methanol(D) states increases monotonously: this can only happen if methanol(D) – water(A) bonds are formed, replacing some of methanol(D) – methanol(A) bonds. Interestingly, the energy of methanol(D) – water(A) pairs is typically higher than that of methanol – methanol pairs, which suggests that the formation of this bond is influenced by the H-bonded environment (multibody-effects).

*Connectivity: Monomers, clusters, cluster size distributions, percolation*

Up to this point, we have considered variations of the hydrogen bonding network from the point of view of single molecules and (H-bonded) molecular pairs. From this point on, the focus is shifted to studying the behavior of the H-bonded network as a whole. It may be expected that the network expands as temperature decreases and larger aggregations, consisting of both compounds, are formed with higher probabilities.

The H-bonded network can be investigated by taking the entire system (all possible kinds of H-bonds, both like and unlike molecules) into account, or for the water and methanol subsystems exclusively, when only H-bonds between like molecules are considered.

First, the behavior of the total system is discussed in the mixtures (pure methanol will be considered while describing the methanol subsystem). The relative frequency of molecules participating in 0, 1, or more H-bonds, as a function of temperature and concentration, can be deducted from Fig. 2 (more details are provided in the SM, in Figs. S6 and S7). The ratio of solitary molecules (monomers), that have no H-bond, is below 3% (for water molecules, it is below 0.1%), and decreases with decreasing temperature and methanol concentration. Molecules with 1 H-bond are the end points of chains, or two of them can form a dimer. The ratio of these molecules among water molecules is also very low, below 1.5%. At high methanol concentrations, a significant amount of methanol molecules have only 1 H-bond (about 22% in $x_M = 0.9$ mixture at 300 K); their occurrence is lower at low temperatures and low methanol concentrations. Most molecules have at least 2 H-bonds; they are the 'interior' molecules in chains and/or cycles.

Molecules connected via H-bonds form H-bonded clusters: if two molecules are connected by a chain of H-bonded molecules then they belong to the same cluster. The size of a cluster ($n_c$) is defined as the number of molecules forming it. (Monomers are clusters with size 1.) The system is said to percolate if the size of the largest cluster is in the order of the number of molecules in the system.

The cluster size distribution for the total system at 300 K is shown in Fig. 8. At low methanol concentrations, the size of the largest cluster is close to the number of molecules in the simulation box (2000). As methanol concentration increases, the largest clusters become smaller. Nevertheless, even for $x_M = 0.9$, their size is around 1600 (80% of the molecules in the simulation box), and the system is still percolating. As temperature decreases, the largest clusters grow significantly; at the lowest temperature considered, their sizes are very close to the system size in methanol-rich mixtures, too (for details, see the SM, Figs. S14 – S16).

The water subsystem (i. e., when only water – water H-bonds are taken into account), quite naturally, behaves differently at low and high methanol concentrations. The first question that may arise is whether there is a, and if so, where is the, percolation transition for water molecules in the mixtures. As methanol concentration decreases below $x_M = 0.6$, the size of water clusters increases rapidly, and 30 (at 300 K) to 50 (at 243 K) % of water molecules form the largest cluster (at $x_M = 0.6$), see Fig. 9. (Note the rather strong dependence on temperature.) In the equimolar mixture, 85 to 93% of water molecules participate in the largest cluster; for $x_M \leq 0.5$, the water subsystem percolates. The percolation threshold is around $x_M = 0.54$ (cf. Figs. 9 and 10): at this methanol concentration the system visibly spans the simulation box at low temperatures (Fig. 10). Upon decreasing the methanol concentration, below $x_M \leq 0.3$, on average, all molecules are connected.

At high methanol concentrations, on the other hand, only small water clusters can be found: at $x_M = 0.9$, the largest water clusters contain just 8 to 10 molecules, and only about 4 to 5% of water molecules can be found in these clusters. The size of water clusters slightly increases as methanol concentration decreases from $x_M = 0.9$ to $x_M = 0.7$, but the ratio of water molecules participating in them remains below 10%.

At the highest methanol concentrations, a significant amount of water molecules have no H-bonded water neighbors at all: at $x_M = 0.9$, more than 60% of water molecules can be considered (water-) monomers. Note that taking into account all types of H-bonds, only a very small (< 0.1%) fraction of water molecules are genuinely monomers; this means that these molecules are completely surrounded by methanol molecules. At this concentration, temperature has only a weak effect on the connectivity of water molecules. As temperature decreases, the number of monomers increases somewhat, and the size of the largest clusters slightly decreases.

The occurrence of monomers in the water subsystem decreases fast with increasing water concentration. At $x_M = 0.6$, only about 10% of water molecules are without an H-bonded water neighbor, and at $x_M \leq 0.4$, in the percolated water sub-network, their ratio falls below 1%. In the mixtures with less than 70 mol% methanol, the number of water monomers decreases with decreasing temperature. (The interested reader may find more details in the SM, see Figs. S17 – S19.)

Turning now to the analysis of methanol clusters, in pure methanol at 300 K, the largest clusters contain only 100 to 150 molecules, whereas at low temperatures, clusters with 1000 to 1200 molecules can also be found, see Fig. 10. In the mixtures, the methanol subsystem is less connected than that of water. The size of the largest cluster, even in pure methanol, only at low temperatures (below 200 K) reaches half of the number of methanol molecules in the simulation box.

The size of the largest methanol cluster decreases with decreasing methanol concentration. For $x_M = 0.9$, the largest methanol clusters contain about 50 – 100 molecules; for $x_M = 0.5$, they consist of about 10 – 20 molecules, and in the $x_M = 0.1$ mixture, only very short ($n_C \leq 5$) methanol chains can be found. Considering the number of methanol molecules in the box, it means that even for $x_M = 0.9$, less than 10% of the molecules constitute the largest cluster. The size of methanol clusters, that increases with decreasing temperature in pure methanol, (essentially) does not depend on the temperature in the mixtures.

The number of methanol monomers is low in pure methanol (below 3.5%) and increases as the methanol concentration decreases. At $x_M \leq 0.8$, at least 10% of methanol molecules have no methanol H-bonded pair. The number of monomers decreases with decreasing temperature at high methanol concentrations ($x_M > 0.8$). At lower methanol concentrations, the abundance of monomers changes only slightly with temperature. (In the $x_M = 0.7$ mixture, it shows a shallow minimum, in the $x_M = 0.5$ mixture it increases with decreasing temperature.) In water-rich mixtures, 40 to 90% of methanol molecules are monomers, i.e., they are surrounded by water molecules.

(We note that in the SM, a more detailed background can be found concerning the above pieces of results on methanol clusters, see Figs. S16, S18, S20, and S21.)

As it was shown in the previous section, the number of H-bonds between methanol molecules increases with decreasing temperature only at high methanol concentrations, similarly to the size of the largest clusters in the methanol subsystem. The number of water – water connections also decreases in methanol-rich mixtures, in parallel with the size of the largest clusters in the water subsystem.

The percolation threshold is around 85 mol% in ethanol – water mixtures, around 89% in 1-propanol – water mixtures, and 62 mol% in 2-propanol – water mixtures ([22 – 24]). In all of the alcohol-water mixtures mentioned above, the percolation threshold increases with decreasing temperature. The water subsystem also percolates in water-rich ethanol – water and 1-propanol – water mixtures.

*Connectivity II: cycles, primitive rings, ring distributions*

Molecules with at least two H-bonds can participate in cycles (cyclic entities), that are H-bonded chains whose last and first molecules are H-bonded to each other. The size of a cycle is defined as the shortest path through all the H-bonded molecules in the cycle. A cycle is called a primitive ring if it cannot be decomposed into smaller rings. Primitive rings up to size 8 are investigated here: they will be referred to as 'rings' hereafter.

The concentration and temperature dependence of the average number of rings ($N_R$) is shown in Fig. 11. At low methanol concentrations, several molecules participate in more than one ring (the number of rings is higher than the number of molecules in the simulation box). As the fraction of methanol molecules increases, the number of rings decreases. In pure methanol, only 3 to 7 rings can be found in the simulation box (containing 2000 methanol molecules). As temperature reduces, the number of rings increases in the water-rich mixtures (as was shown earlier [34]); it has a maximum in the methanol-rich mixtures, and it decreases in pure methanol (see Fig. 11, as well as background material in the SM).

The distribution of ring sizes is provided by Fig. 12. It was found earlier [34] that in mixtures with $0.2 \leq x_M \leq 0.4$, 6-membered rings become the most frequent as the temperature approaches the freezing point. Our simulations yielded similar results at low methanol concentrations ($x_M \leq 0.3$). Six-membered rings are the most common even at room temperature at the lowest methanol concentration ($x_M = 0.1$). For

$x_M \geq 0.2$, five-membered rings are the most frequent at 300 K. In the water-rich mixtures, the number of 5, 6, and 7-membered rings increases with decreasing temperature. At the lowest investigated temperature, six-membered rings are the most frequent for $x_M$ = 0.1, 0.2, and 0.3, and they are similarly prevalent as the five-membered rings for $x_M$ = 0.4 and 0.5. In the methanol-rich mixtures, the number of rings (after a short increase) decreases as temperature decreases. Larger rings are preferred at the lowest investigated temperatures in the methanol-rich mixtures.

Rings can contain only one type (water rings and methanol rings), or both types of molecules (mixed rings). The ring type distribution at 300 K is shown in Fig. 13. Nearly 80% of the rings are exclusively water rings in the $x_M$ = 0.1 mixture. The ratio of mixed rings increases rapidly as methanol concentration increases; this ring type dominates in mixtures with methanol concentration $x_M \geq 0.3$. The number of methanol rings is very low at all concentrations: only 2 to 3 methanol rings can be found in the $x_M$ = 0.9 mixture. The ratio of different ring types shows only weak temperature dependence: the relative frequencies of mixed rings increase slightly with decreasing temperature. The number of rings follows the temperature dependence of the abundance of mixed rings. The number of purely water rings decreases more strongly in methanol-rich mixtures than the number of mixed rings does. (Further background information is available in the SM.)

Finally, it is worth noting that the preference of 6-membered rings is a unique property of methanol – water mixtures. Similar behavior has been reported for other alcohol – water mixtures only at very low alcohol concentrations ($x_A \leq 0.1$). At higher alcohol concentrations, 5-membered rings are the most popular in ethanol – water, 1-propanol – water, and 2-propanol – water mixtures [22 – 24].

**Conclusions**

Methanol – water mixtures have been investigated by classical molecular dynamics simulations over the entire concentration range (at methanol molar fractions 0.1 to 1.0) at room temperature and down to the freezing point of the mixtures. The OPLS-AA force field for methanol and the TIP4P/2005 model of water were applied in the simulations. The hydrogen-bonded network was analyzed, based on the determination of the number of H-bonds between different types of molecules.

It has been found that upon cooling:

- The average H-bond number of water molecules increases at every concentration and temperature investigated.
- The average H-bond number of methanol molecules increases at low ($x_M \leq 0.4$) and high ($x_M \geq 0.9$) methanol concentrations, while it approaches a maximum value between 180 and 200 K in the concentration region $0.5 \leq x_M \leq 0.8$.

- The average number of H-bonds between unlike molecules increases over the entire concentration and temperature region. However, the magnitude of the increase is different for methanol(D) – water(A) and water(D) – methanol(A) pairs, and it is highest for M(D) – W(A).
- The number of H-bonds between water molecules increases in water-rich, while it has a maximum or decreases in methanol-rich mixtures.
- The number of H-bonds between methanol molecules increases for $x_M \geq 0.9$, whereas it has a maximum, or it even decreases, over the entire temperature range at concentrations $x_M \leq 0.8$.
- The numbers of both methanol and water molecules acting as H-bond donors approach their theoretical maximum value.
- The number of water molecules as H-bond acceptors increases monotonously, whereas it reaches a maximum value for methanol molecules at intermediate concentrations.

Regarding the connectivity of the H-bonded network, we were able to make the following observations:
- The H-bonded network percolates in each mixture: the percolation threshold is above $x_M = 0.9$ at room temperature.
- The water subsystem percolates in mixtures with $x_M \leq 0.5$. The methanol subsystem forms a percolating network only in (very) low-temperature pure methanol.
- Water molecules like to form rings: the number of primitive rings increases upon cooling in water-rich mixtures. Six-membered rings are the most common for mixtures where $x_M \leq 0.3$.
- Most rings contain both water and methanol molecules (mixed rings). Water-only rings are the most frequent in mixtures with $x_M \leq 0.2$.

**Acknowledgment**

The authors acknowledge financial support from the National Research, Development, and Innovation Office (NKFIH) under grant No. K142429. Financial support from the Eötvös Loránd Research Network (ELKH, Hungary) via their special fund, Grant No. SA-89/2021. is also acknowledged. We thank Dr. Imre Bakó for sharing the computer code used for hydrogen bond analysis.

**Table 1** Simulation stages following the initial energy minimization. In every step, the leap-frog algorithm was applied with a time step of 2 fs. ($T_{inv}$ denotes the actual temperature.)

|  | Temperature [K] | Ensemble | run time [ns] | Thermostat / time constant [ps] | Barostat / time constant [ps] |
|---|---|---|---|---|---|
| High temperature equilibration | 320 K | NVT | 0.2 | Berendsen [50] / 0.1 | - |
| Heat treatment | 320 K | NpT | 2 | Nose-Hoover [51,52] / 0.5 | Parrinello-Rahman [53,54] / 2.0 |
| Cooling down | 320 K → $T_{inv}$ | NpT | 5 ns / K | Berendsen / 0.1 | Berendsen [50] / 0.3 |
| Equilibration | $T_{inv}$ | NpT | 10 | Nose-Hoover / 0.5 | Parrinello-Rahman / 2.0 |
| Equilibration | $T_{inv}$ | NVT | 2 | Nose-Hoover / 0.5 | - |
| Production | $T_{inv}$ | NVT | 20 | Nose-Hoover / 0.5 | - |

**Table 2** Methanol – water mixtures and temperatures considered here, together with the lengths of the simulation boxes (in nm) at the given temperatures.

| $x_M$ | 300 K | 268 K | 263 K | 253 K | 243 K | 233 K | 223 K | 213 K | 203 K | 193 K | 178 K | 163 K |
|---|---|---|---|---|---|---|---|---|---|---|---|---|
| 0.1 | 4.05083 | 4.04519 | 4.02668 | | | | | | | | | |
| 0.2 | 4.18204 | 4.17505 | - | 4.15342 | 4.15237 | | | | | | | |
| 0.3 | 4.31986 | 4.29243 | - | 4.27225 | 4.25668 | 4.25969 | 4.24066 | | | | | |
| 0.4 | 4.45790 | 4.40894 | - | - | 4.38098 | 4.35825 | 4.36251 | 4.34207 | | | | |
| 0.5 | 4.57034 | 4.52553 | - | - | 4.47417 | 4.47759 | - | 4.44539 | 4.44042 | 4.41984 | | |
| 0.5442 | 4.60581 | 4.55462 | - | - | - | 4.51702 | - | - | - | 4.46669 | | |
| 0.6 | 4.68123 | 4.63225 | - | - | 4.59623 | 4.57495 | - | 4.55567 | - | 4.52415 | 4.50793 | |
| 0.7 | 4.82393 | 4.73562 | - | - | - | 4.68131 | - | - | 4.64357 | 4.63029 | 4.60168 | 4.59020 |
| 0.7337 | 4.83865 | 4.79665 | - | - | - | 4.73329 | - | - | - | 4.66948 | 4.63855 | 4.61774 |
| 0.8 | 4.92104 | 4.87076 | - | - | - | 4.80481 | - | - | 4.74596 | 4.72972 | 4.70644 | 4.67427 |
| 0.9 | 5.04533 | 4.97809 | - | - | - | 4.90044 | - | - | - | 4.83031 | 4.80166 | 4.77417 |
| 1.0 | 5.17028 | 5.09110 | - | - | - | 5.01612 | - | - | 4.93355 | 4.92606 | 4.89494 | 4.86484 |

**Figures**

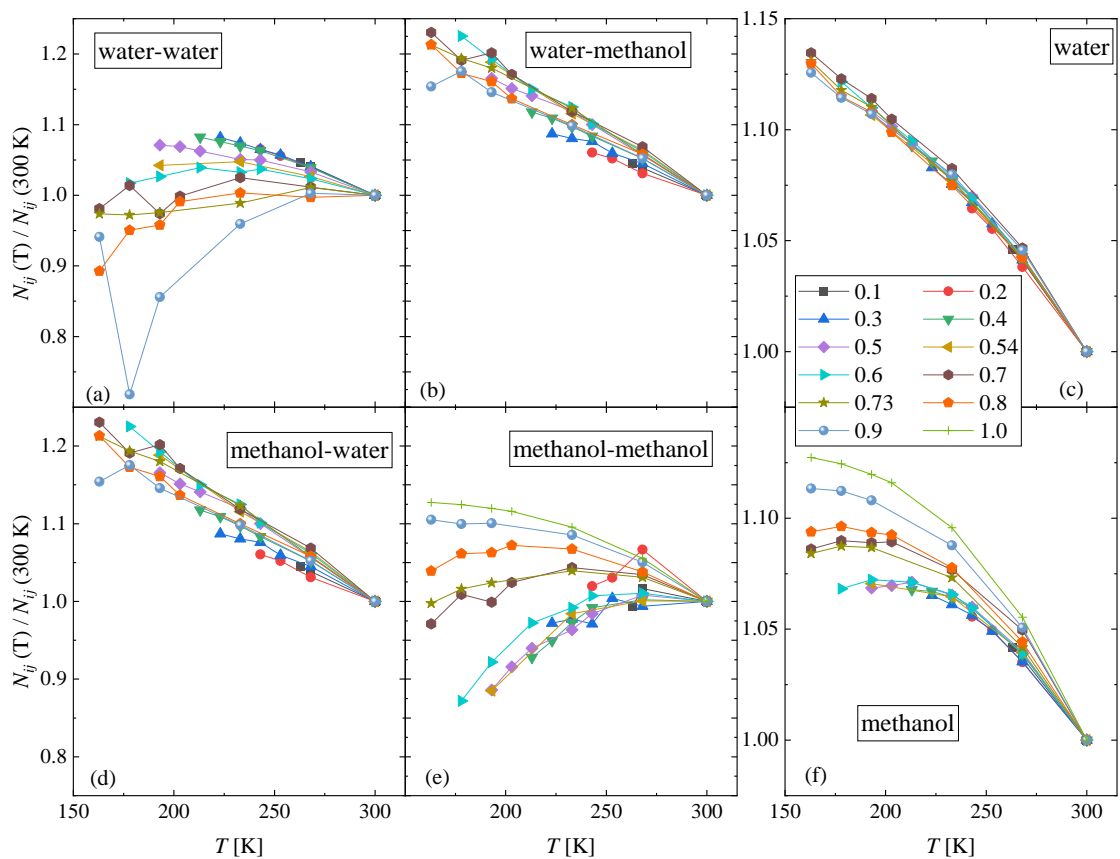

**Figure 1.** Temperature dependence of the number of hydrogen bonds at different concentrations, normalized to the 300 K values: (a) the average number of H-bonded water molecules around water, (b) the average number of H-bonded methanol molecules around water, (c) the average number of H-bonded (water and methanol) molecules around water, (d) the average number of H-bonded water molecules around methanol, (e) the average number of H-bonded methanol molecules around methanol, (f) the average number of H-bonded (water and methanol) molecules around methanol.

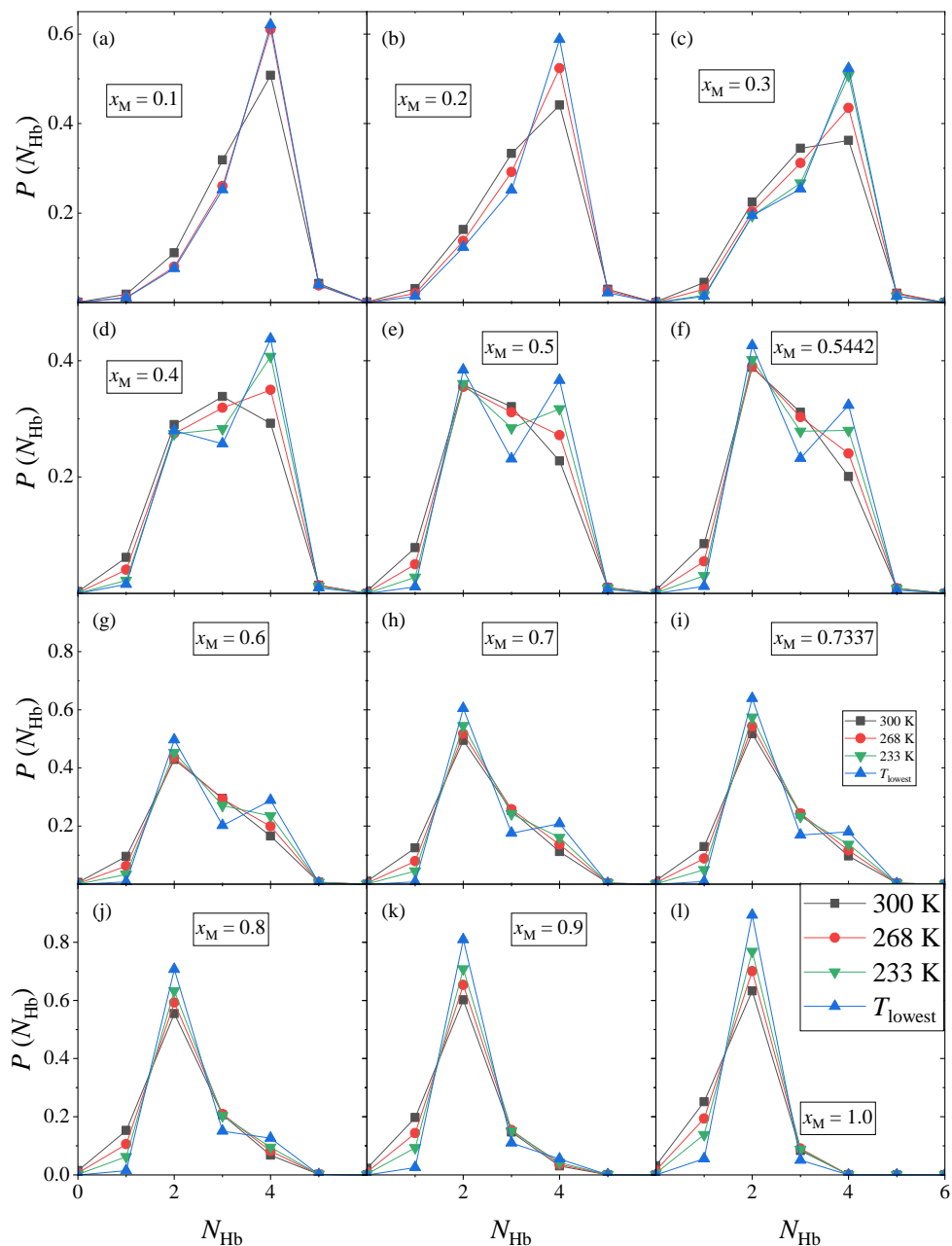

**Figure 2.** Distribution of the number of H-bonds at different temperatures and methanol concentrations in methanol-water mixtures. The $T_{\text{lowest}}$ temperature (the lowest investigated temperature at a given methanol concentration) is the following: (a) $x_M = 0.1$, 263 K; (b) $x_M = 0.2$, 243 K; (c) $x_M = 0.3$, 223 K; (d) $x_M = 0.4$, 213 K; (e, f) $x_M = 0.5$, and 0.5442, 193 K; (g) $x_M = 0.6$, 178 K; (h) – (l) $0.7 \leq x_M \leq 0.9$, 163 K. (The lines are just guides to the eye.)

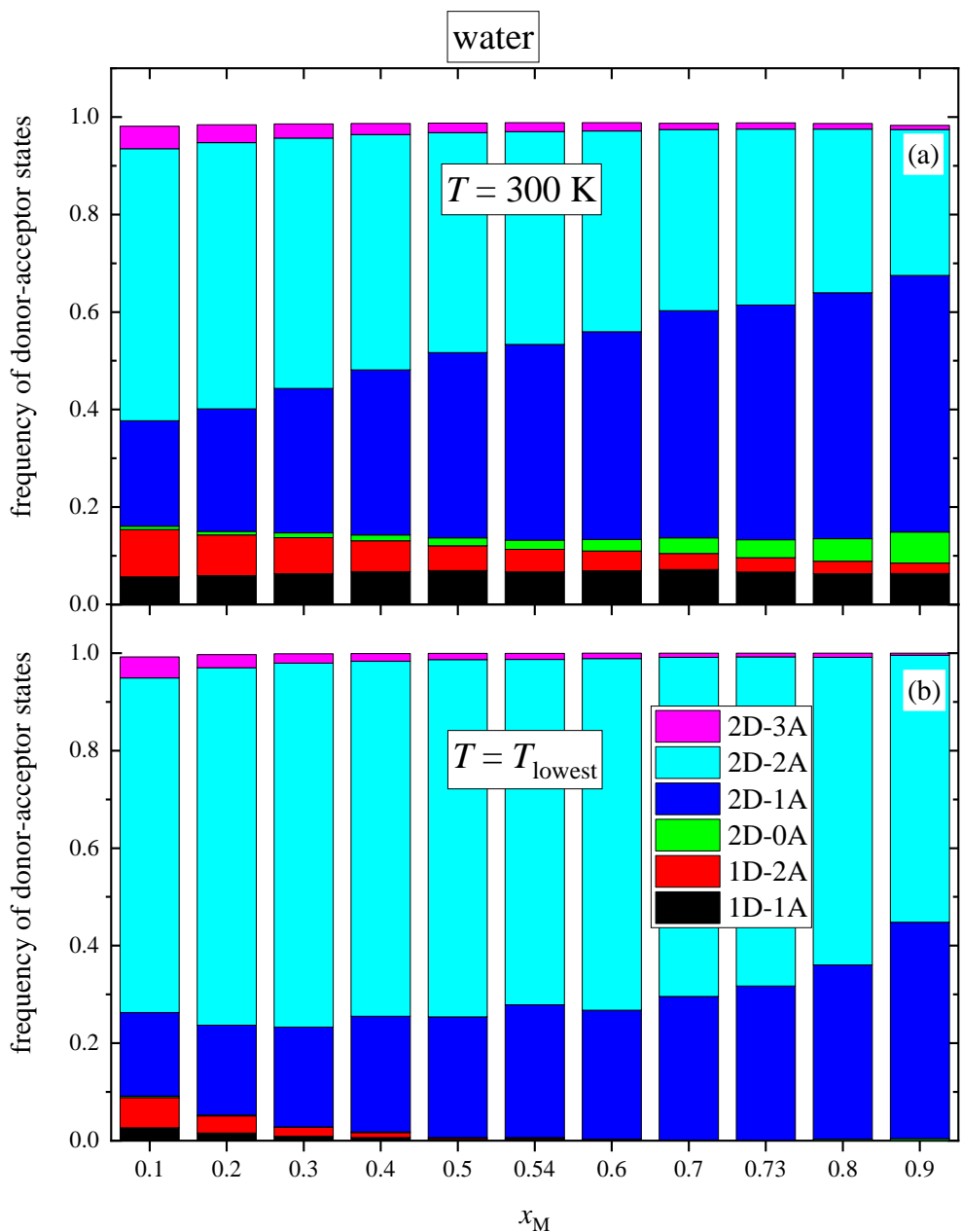

**Figure 3.** Relative frequencies of the most frequent donor-acceptor states of water molecules in methanol – water mixtures (a) at 300 K and (b) at the lowest investigated temperature ($T_{\text{lowest}}$ = 263 K for 0.1, 243 K for 0.2, 223 K for 0.3, 213 K for 0.4, 193 K for 0.5 and 0.54, 178 K for 0.6 and 163 K for $x_M \geq 0.7$ mixtures).

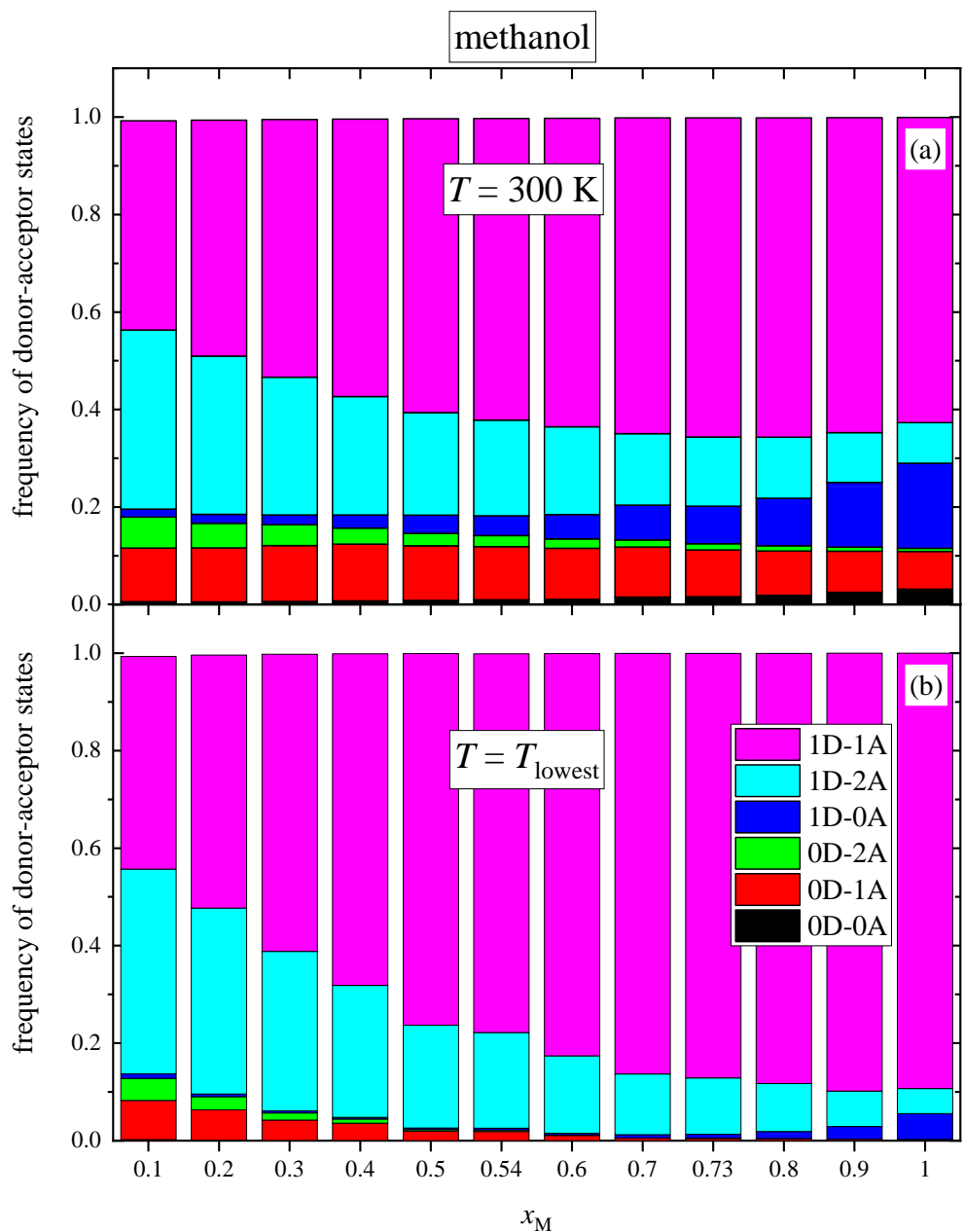

**Figure 4.** Relative frequencies of the most frequent donor-acceptor states of methanol molecules in methanol – water mixtures (a) at 300 K and (b) at the lowest investigated temperature ($T_{\text{lowest}} = 263$ K for 0.1, 243 K for 0.2, 223 K for 0.3, 213 K for 0.4, 193 K for 0.5 and 0.54, 178 K for 0.6 and 163 K for $x_M \geq 0.7$ mixtures).

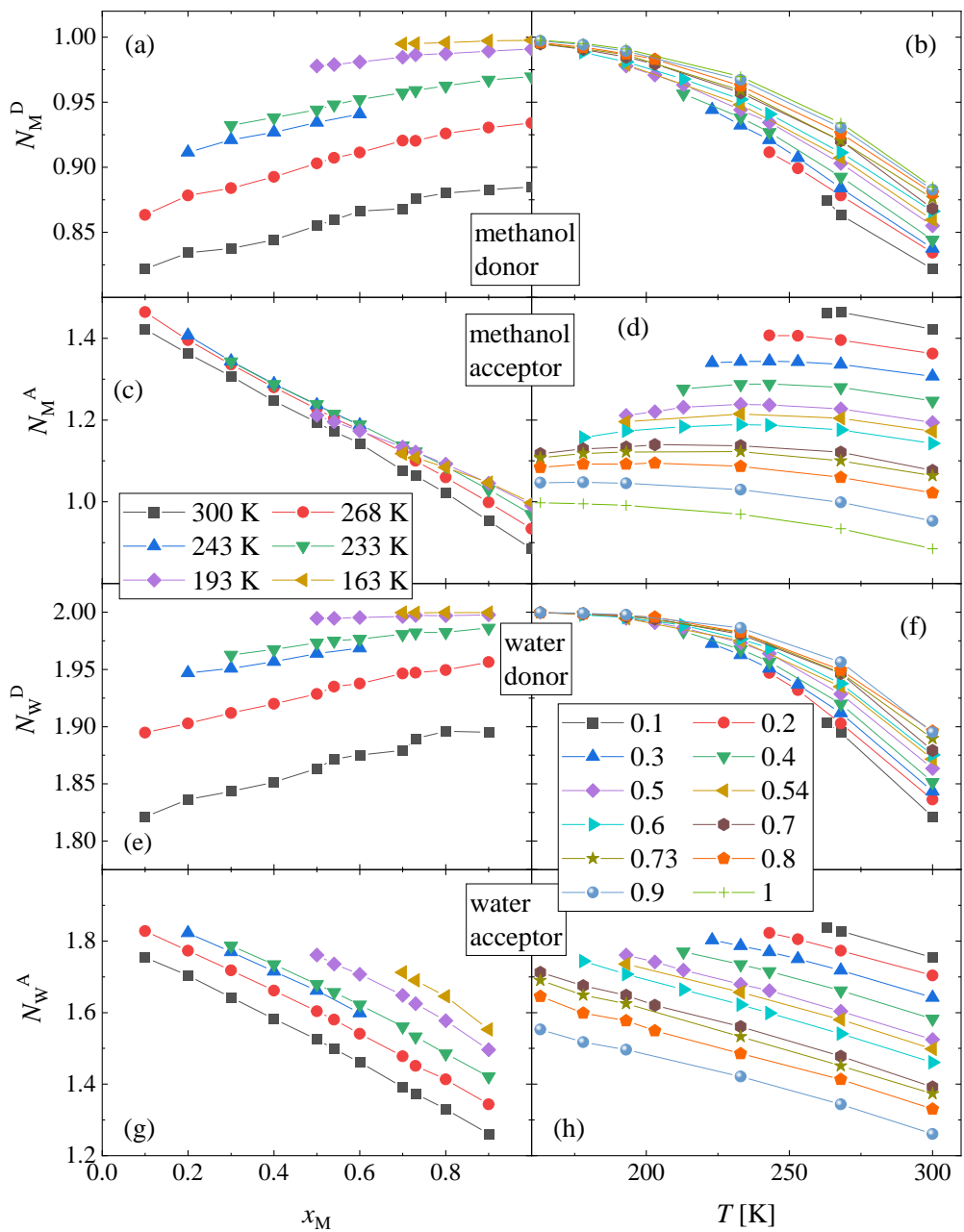

**Figure 5.** (left side) Concentration and (right side) temperature dependence of the average number of H-bonds of (a-d) methanol and (e-h) water molecules as (a, b, e, f,) donors and as (c, d, g, h) acceptors.

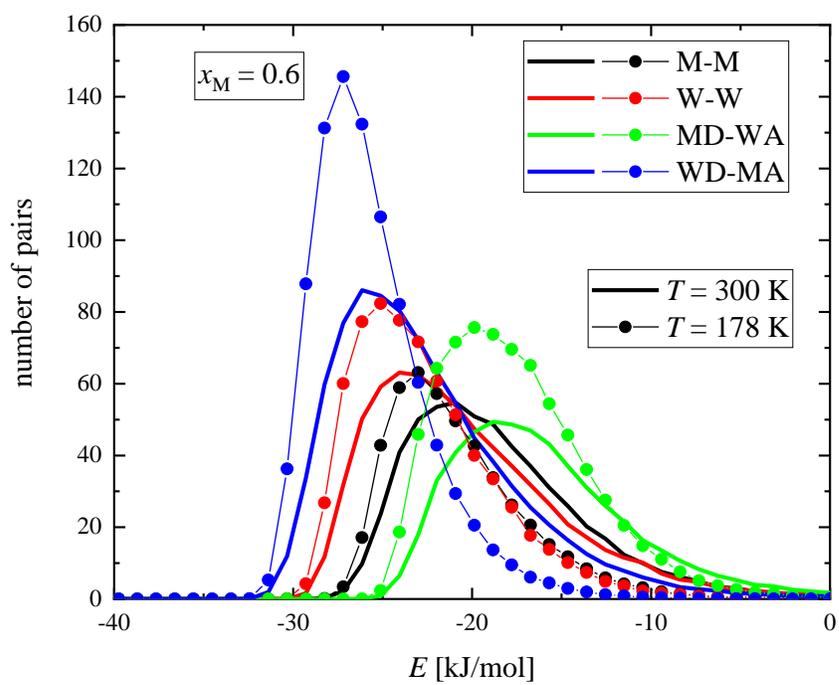

**Figure 6.** Energy distributions of the different H-bonded pairs in methanol – water mixtures for $x_M = 0.6$ at (thick lines) 300 K and (symbols with thin lines) at 163 K.

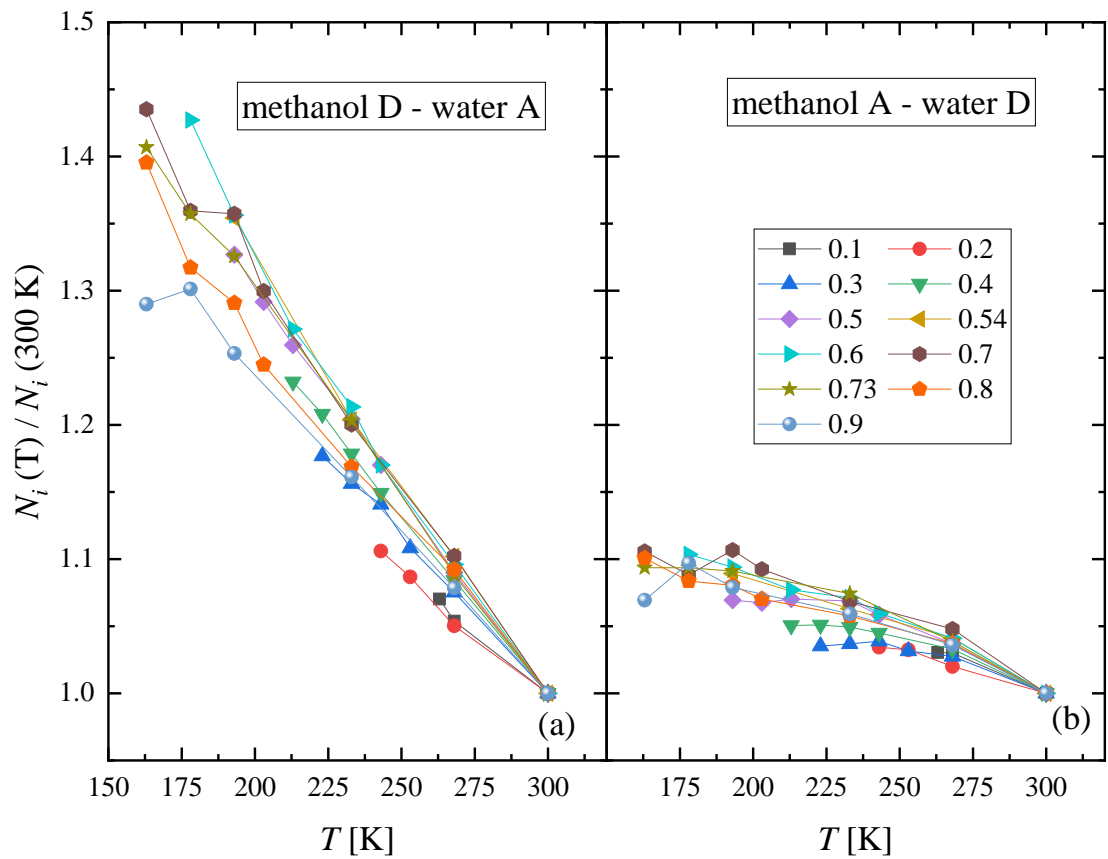

**Figure 7.** Temperature dependence of the average number of (a) methanol(D) – water(A), and (b) methanol(A) – water(D) H-bonds, normalized to the 300 K values, at different methanol concentrations.

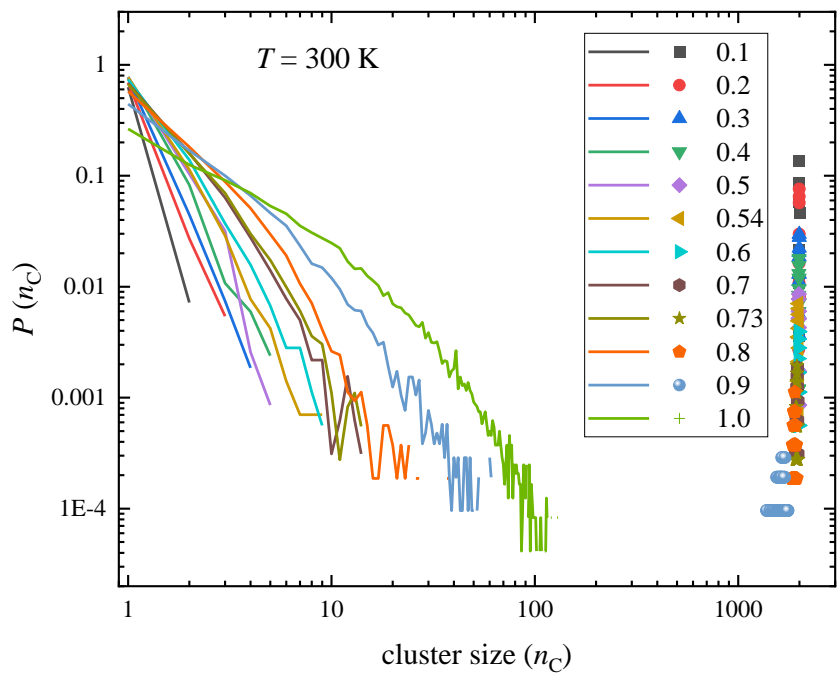

**Figure 8.** Cluster size distributions in methanol-water mixtures at $T = 300$ K, considering H-bonds between any types of molecules.

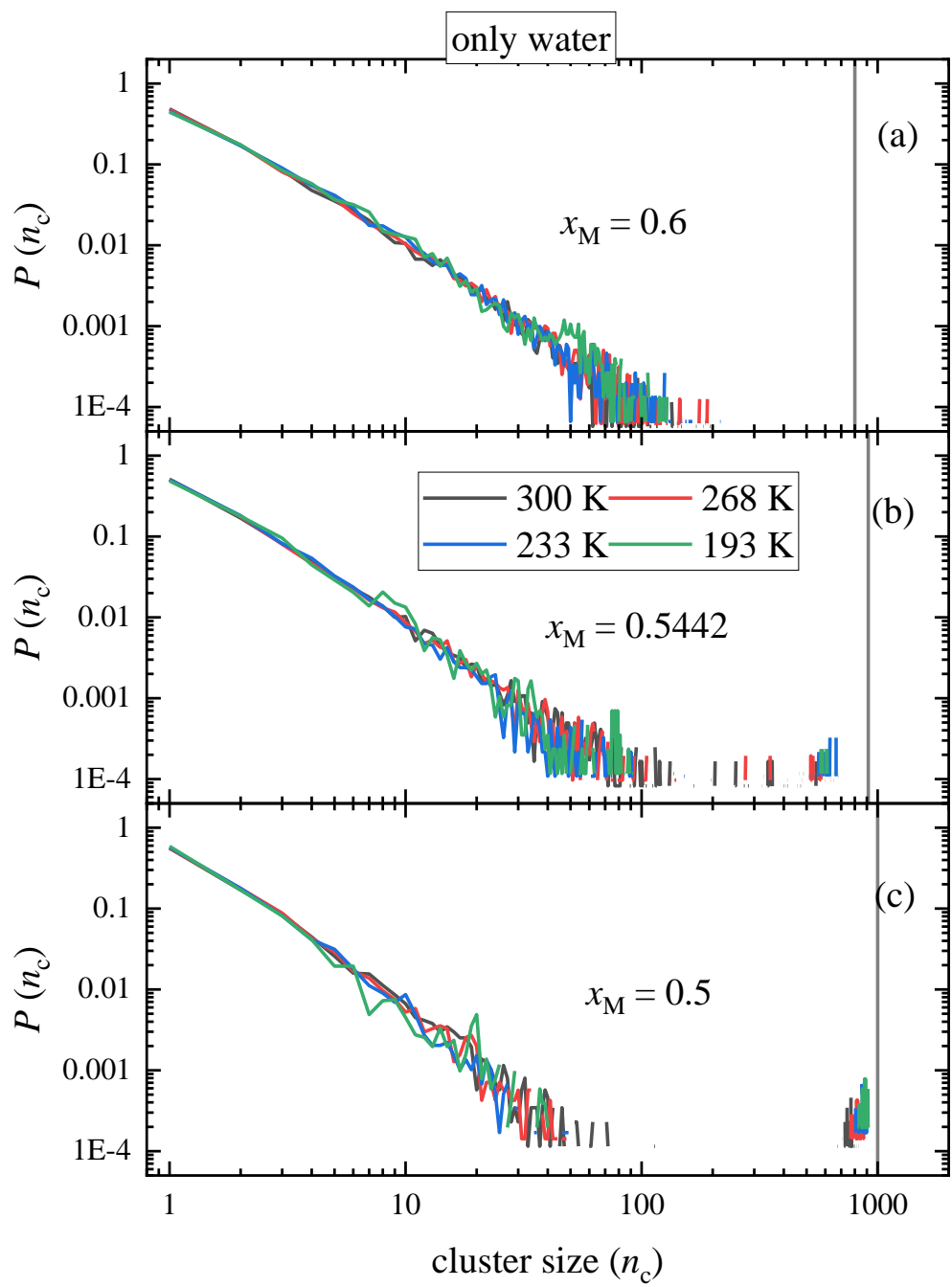

**Figure 9.** Cluster size distributions within the water subsystem in methanol – water mixtures at intermediate concentrations at different temperatures (H-bonds only between water molecules are considered.) The vertical lines denote the numbers of water molecules in the simulation boxes.

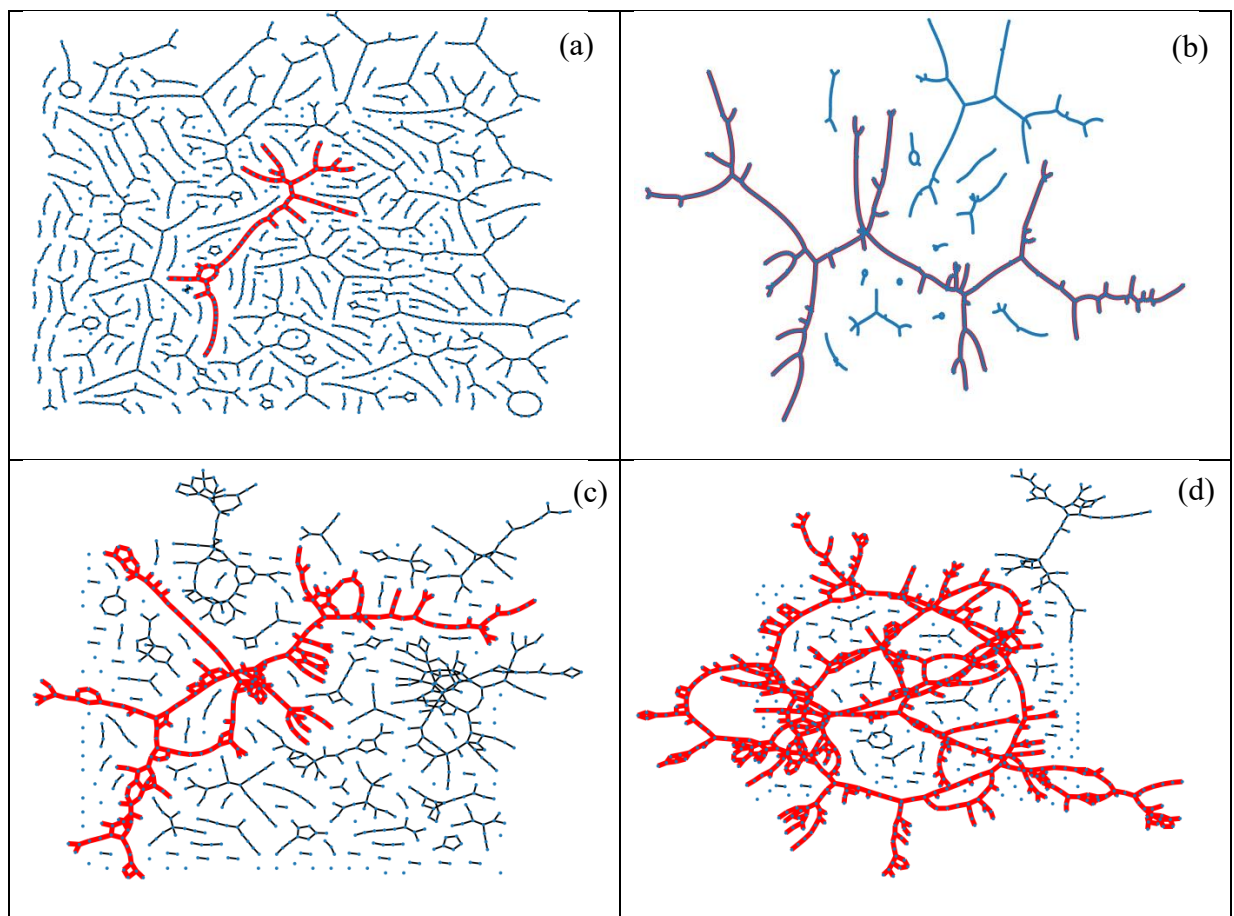

**Figure 10.** Typical H-bond topology of (a, b) pure methanol (a) at $T = 300$ K, and (b) at $T = 163$ K, and (c, d) of the water – subsystem in the methanol – water mixture of $x_M = 0.54$ (c) at $T = 300$ K, and (d) $T = 268$ K. Blue dots: molecules; black lines: H-bonds; red thick lines: H-bonds belong to the largest cluster.

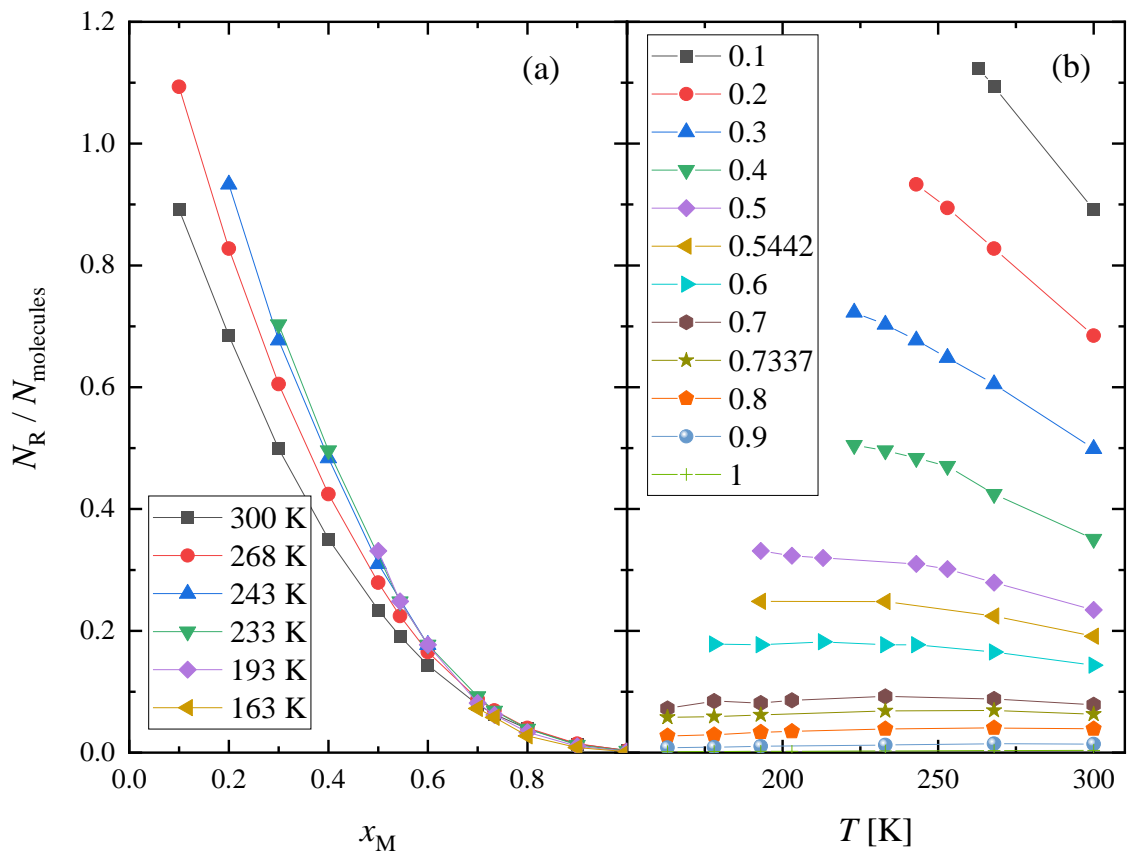

**Figure 11.** (a) Concentration and (b) temperature dependence of the average number of primitive rings per molecule.

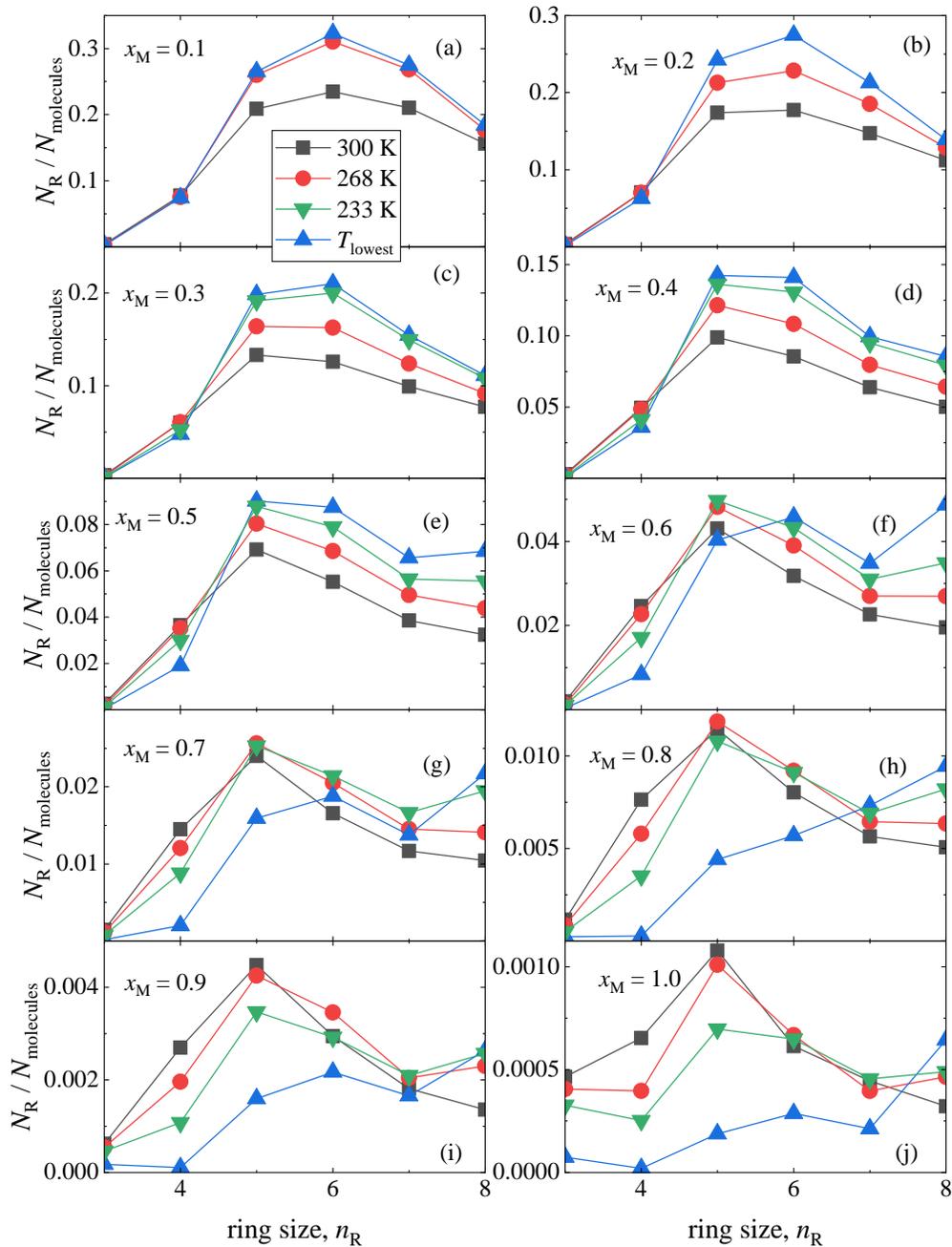

**Figure 12.** Ring size distributions, normalized by the number of molecules in methanol-water mixtures at different temperatures. The $T_{lowest}$ temperatures (the lowest investigated temperature at a given methanol concentration) are the following: (a) $x_M = 0.1$, 263 K; (b) $x_M = 0.2$, 243 K; (c) $x_M = 0.3$, 223 K; (d) $x_M = 0.4$, 213 K; (e) $x_M = 0.5$, 193 K; (f) $x_M = 0.6$, 178 K; (g) – (j) $0.7 \leq x_M \leq 1.0$, 163 K. (Connecting lines are just guides to the eye.)

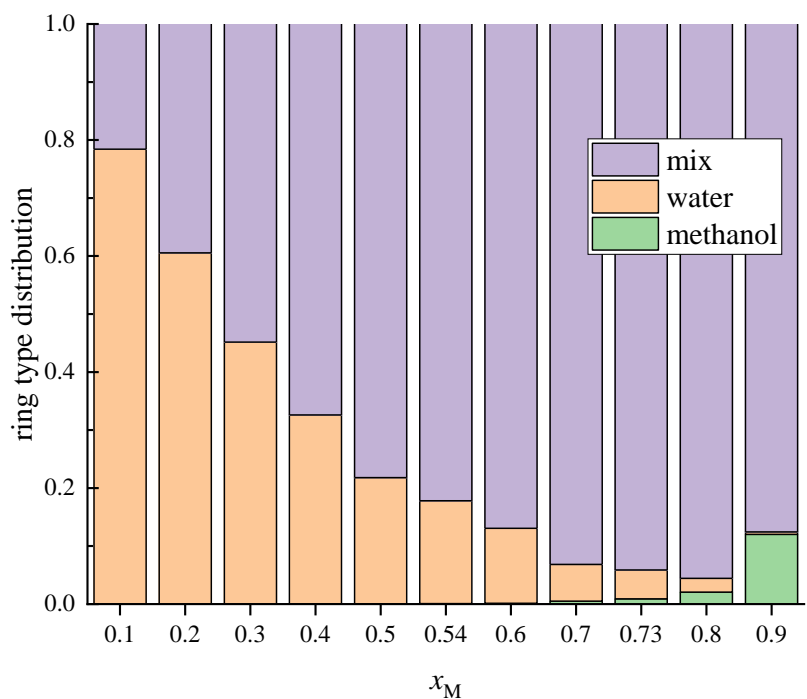

**Figure 13.** Frequencies of different ring types (rings containing only water, only methanol or both water and methanol molecules) in methanol – water mixtures at 300 K.

# Supplementary Material

# Evolution of the hydrogen-bonded network in methanol-water mixtures upon cooling

Ildikó Pethes[1,a], László Pusztai[a,b], László Temleitner[a]

[a]Wigner Research Centre for Physics, Konkoly Thege út 29-33., H-1121 Budapest, Hungary

[b]International Research Organization for Advanced Science and Technology (IROAST), Kumamoto University, 2-39-1 Kurokami, Chuo-ku, Kumamoto 860-8555, Japan

---

[1]Corresponding author: e-mail: pethes.ildiko@wigner.hu

*Number of H-bonds*

The average number of H-bonds per molecule for the total system (methanol and water molecules together, $N_{Hb}$), for the two kinds of molecules separately ($N_M$ and $N_W$, for methanol and water molecules, respectively), and for the different pairs of molecules ($N_{MM}$, $N_{MW}$, $N_{WW}$, and $N_{WM}$, for methanol molecules around methanol, water molecules around methanol, water molecules around water, and methanol molecules around water, respectively) are shown in Figs. S1–S3, as the function of temperature and methanol content. What is new in this work is that a kind of 'heat treatment' was used for avoiding the aggregation of molecules, along with 'simulated annealing' for obtaining more reasonable configurations at low temperatures and high alcohol concentrations (see section 2 of the main text). The H-bond numbers obtained by this method are mostly the same as those in our previous work[S1], with slight differences only in the region mentioned before; these slight deviations are an indication of the uncertainties of the calculations.

The average H-bond number ($N_{Hb}$) and the average number of H-bonded molecules around water molecules ($N_W$) increase monotonously with decreasing temperature, see Figs. S4, and Fig 1 of the main text, where the numbers are normalized by the 300 K values.

*H-bond number distributions*

H-bond number distributions of methanol and water molecules separately are shown in Fig. S5 for some concentrations. A detailed account of the concentration and temperature dependence of the numbers of water and methanol molecules with different numbers of H-bonds is presented in Figs. S6 and S7.

*Donor-acceptor states*

The temperature and concentration dependence of the most popular donor – acceptor states of water and methanol molecules is shown in Figs. S8 and S9.

The ratio of the donor and acceptor H-bonds ($N_W^D/N_W^A$, and $N_M^D/N_M^A$) is shown in Fig. S10. The ratios of the average number of donor and acceptor H-bonds of methanol and water molecules (Fig. S10) also reflect the different behavior of water and methanol molecules. As the temperature decreases the $N_W^D/N_W^A$ ratio (which is always higher than 1) decreases. The $N_M^D/N_M^A$ ratio (which is always less than 1) increases with decreasing temperature.

The temperature and concentration dependencies of the average number of H-bonded A and D methanol molecules around water (per water molecule, $N_{WD-MA}$, and $N_{WA-MD}$) are shown in Fig. S11. The average number of H-bonded A and D water molecules around methanol (per methanol molecule, $N_{\text{MA-WD}}$, and $N_{\text{MD-WA}}$) are shown in Fig. S12. For completeness, the concentration and temperature dependence of the ratio of these pairs is shown in Fig. S13.

*Connectivity: Monomers, clusters, cluster size distributions, percolation*

The temperature dependence of the cluster size distributions is demonstrated in Figs. S14 and S15 for the entire system (i.e., all H-bonds are counted).

The temperature and concentration dependence of the size of the largest cluster is shown in Fig. S16 considering the total system and both of the subsystems as well.

The cluster size distributions for the water subsystem are shown in Figs. S17 and S19, and for the methanol subsystem are presented in Figs. S20 and S21.

The concentration and temperature dependence of the average number of monomer molecules considering the two subsystems is presented in Fig. S18.

*Connectivity II: cycles, primitive rings, ring distributions*

The ratio of molecules that participate in cycles and molecules that participate in any chains (i.e., molecules that have at least two H-bonds) is shown in Fig. S22. (Note that molecules may belong to more than one cycle.) Water molecules prefer to organize themselves into cycles: in water-rich mixtures, nearly all molecules with ≥ 2 H-bonds participate in cyclic entities. The ratio of molecules in cycles decreases as methanol concentration increases. In pure methanol, the molecules are in chains and the fraction of molecules-in-cycles is very small. The ratio of molecules in cycles increases with decreasing temperature. This effect is the most pronounced around $x_M = 0.9$: in this mixture, about 50% of molecules with ≥ 2 H-bonds participate in cycles at 300 K, and more than 90% are cycle members at 163 K.

The temperature and concentration dependence of the ring type distribution can be seen in Fig. S24. The ratio of different ring types shows only weak temperature dependence: the relative frequencies of mixed rings increase slightly with decreasing temperature. The number of rings follows the temperature dependence of the abundance of mixed rings. The number of purely water rings decreases more strongly in methanol-rich mixtures than the number of mixed rings does.

Rings can be further categorized according to their composition: if they contain $R_M$ methanol and $R_W$ water molecules then they are denoted as called as '$R_M - R_W$' rings. The concentration and temperature dependence of the ring type distribution is shown in Fig. S25. At low methanol concentrations, rings with 0 or 1 methanol are the most frequent (0 – 5, 0 – 6, 0 – 7). The number of these rings increases as temperature decreases. In the $0.4 \leq x_M \leq 0.6$ mixtures, the 1 – 4 type ring is the most popular at 300 K. At higher methanol concentrations, rings with larger numbers of methanol molecules (2 – 3, 3 – 2, 4 – 1) are the most frequent at room temperature. In the $x_M \geq 0.6$ mixtures, the number of larger rings increases the most with decreasing temperature: the 3 – 4 and 4 – 4 rings are the most abundant at the lowest investigated temperature.

**Figures**

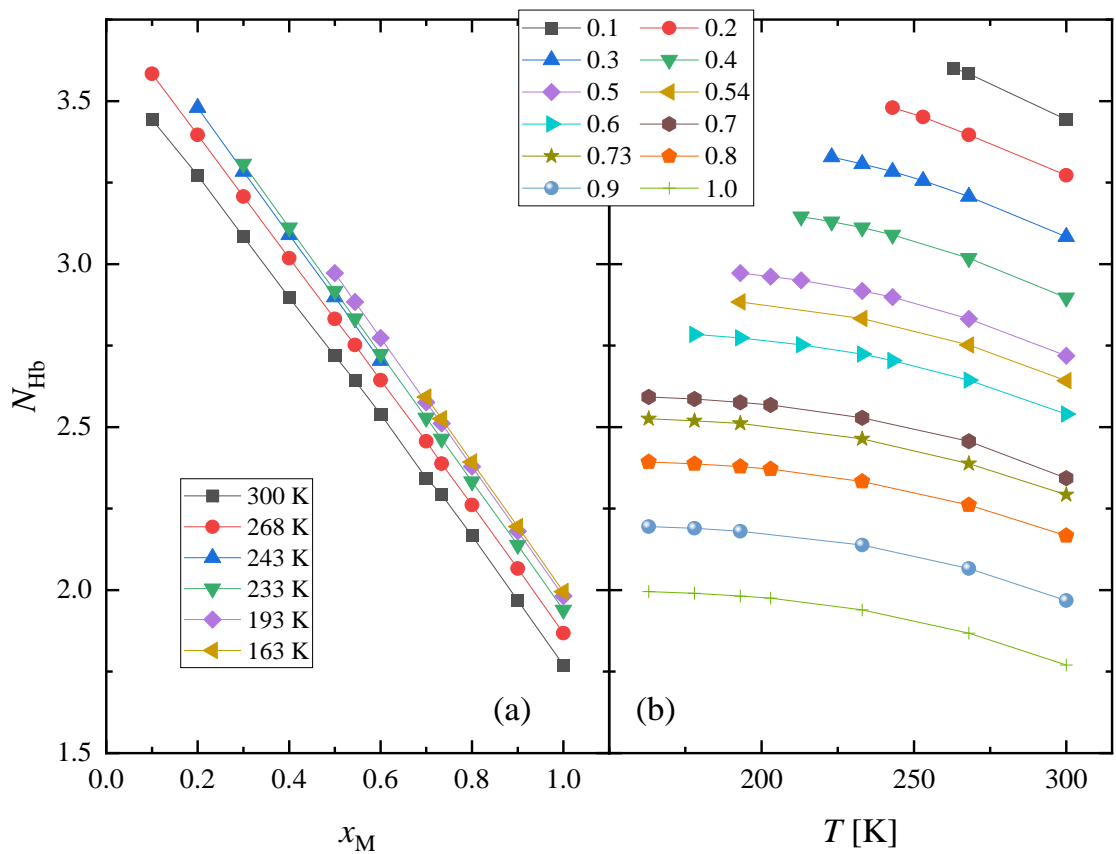

**Figure S1**. The average number of hydrogen bonds per molecule in methanol – water mixtures as a function of (a) temperature at different concentrations and (b) concentration at different temperatures, as obtained by MD simulations.

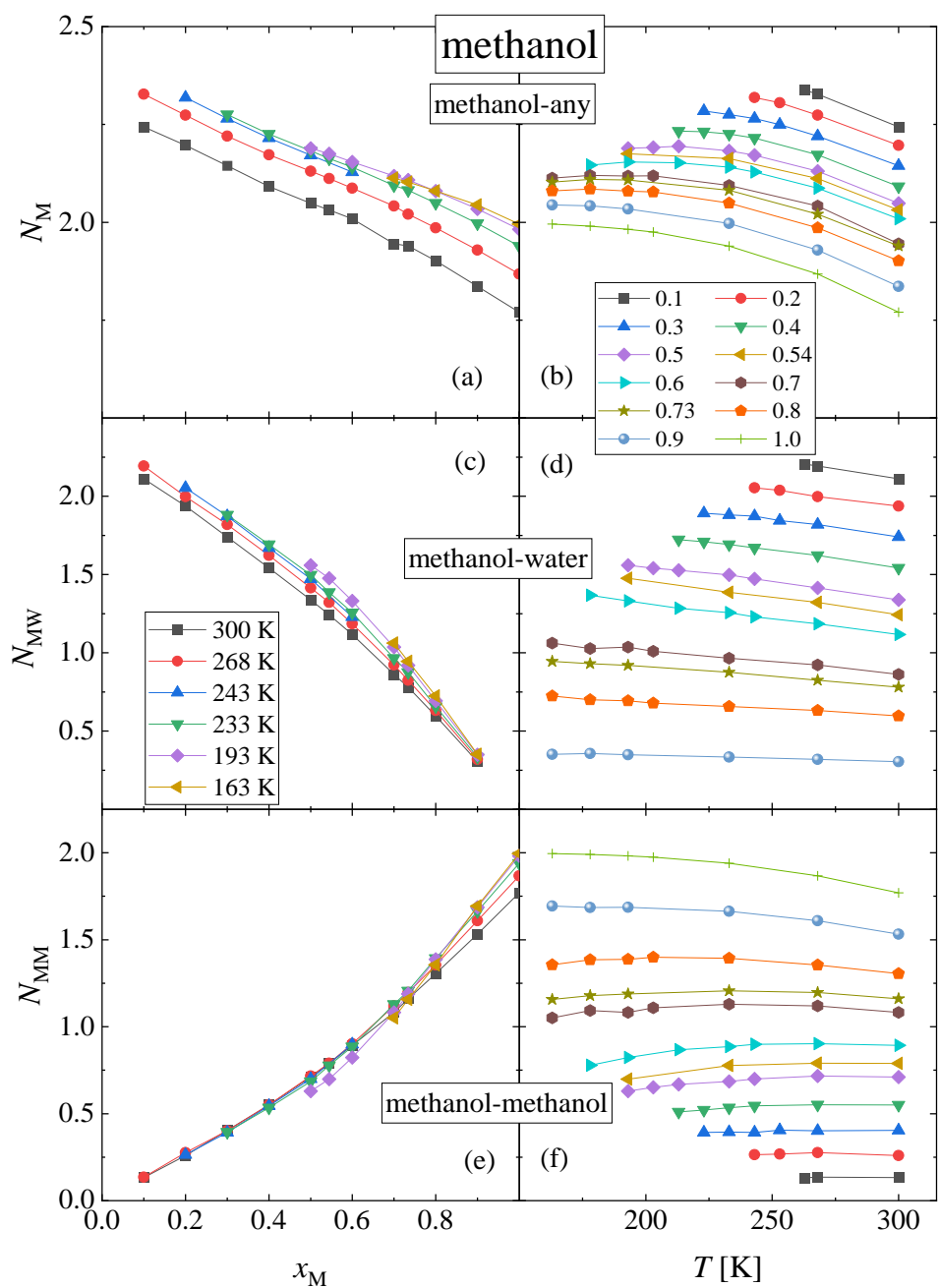

**Figure S2.** (Left panels) concentration and (right panels) temperature dependence of (a and b) the average number of H-bonded (water and methanol) molecules around methanol, (c and d) the average number of H-bonded water molecules around methanol, (e and f) the average number of H-bonded methanol molecules around methanol.

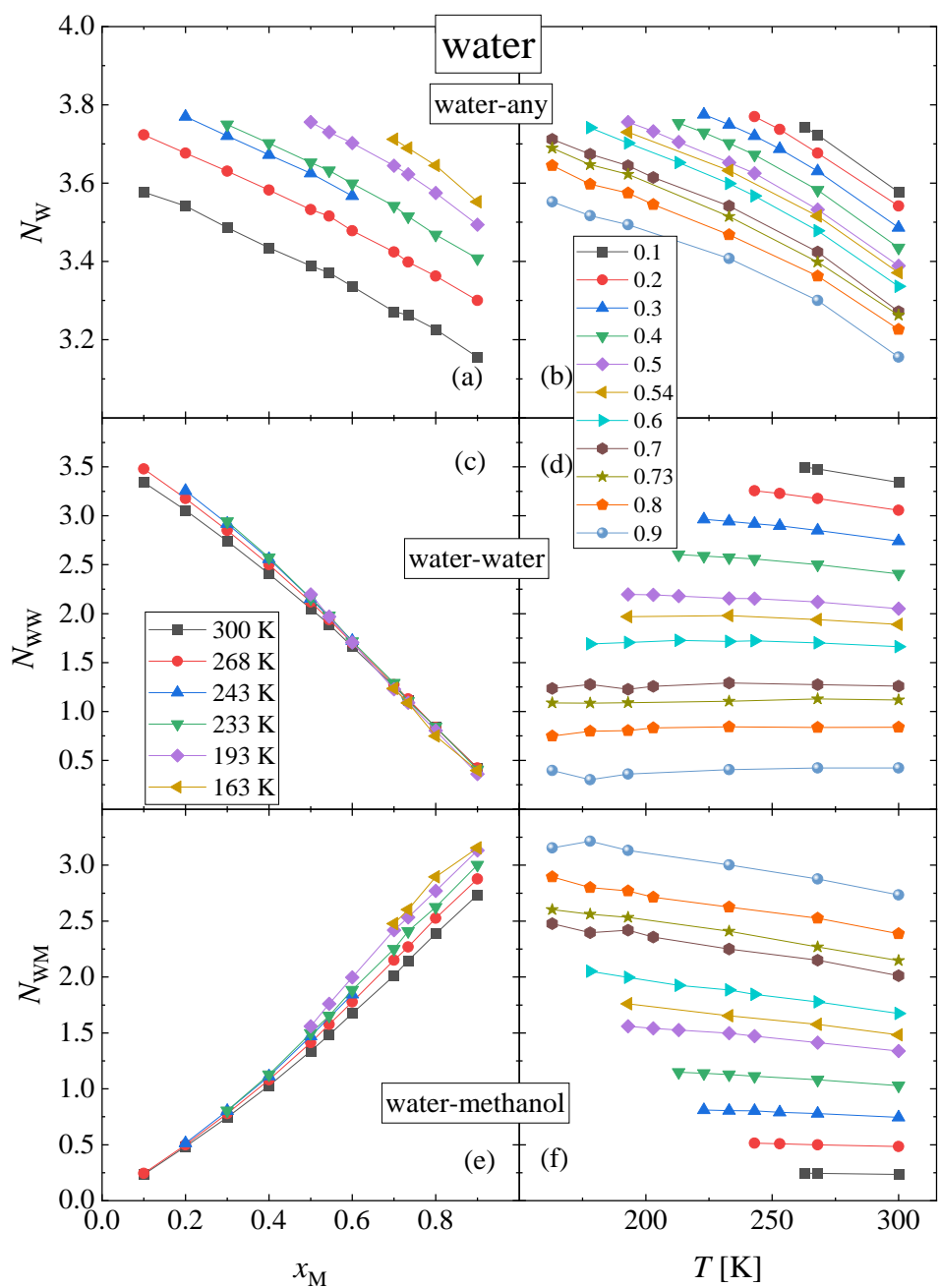

**Figure S3.** (Left panels) concentration and (right panels) temperature dependence of (a and b) the average number of H-bonded (water and methanol) molecules around water, (c and d) the average number of H-bonded water molecules around water, (e and f) the average number of H-bonded methanol molecules around water.

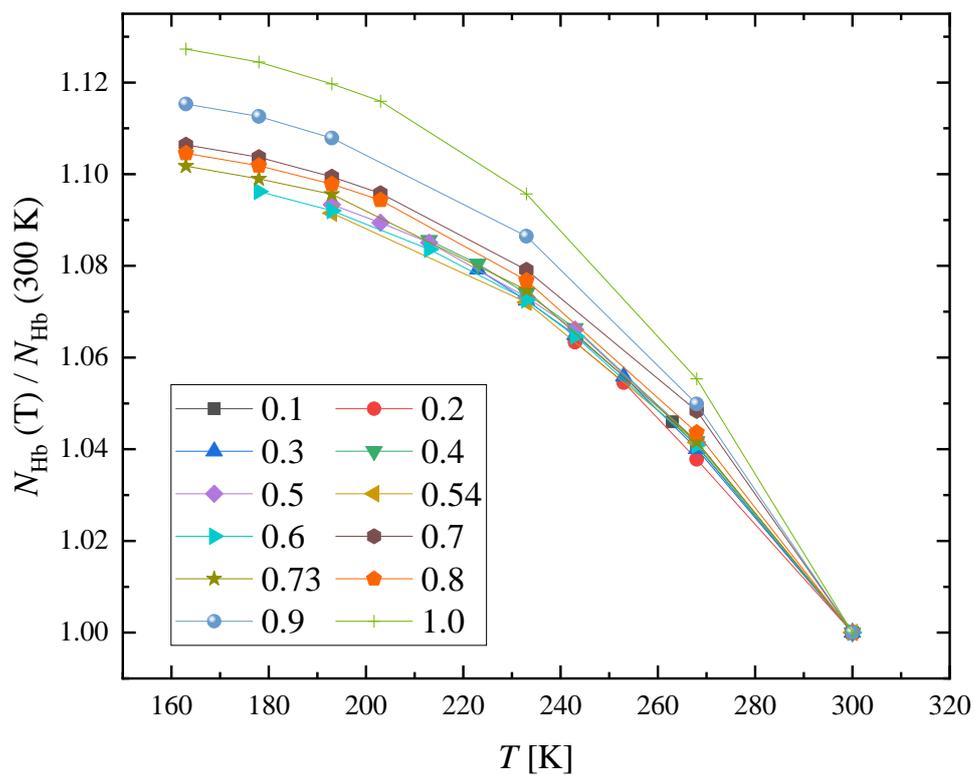

**Figure S4.** Temperature dependence of the average number of hydrogen bonds per molecule, normalized to the 300 K values at different methanol concentrations.

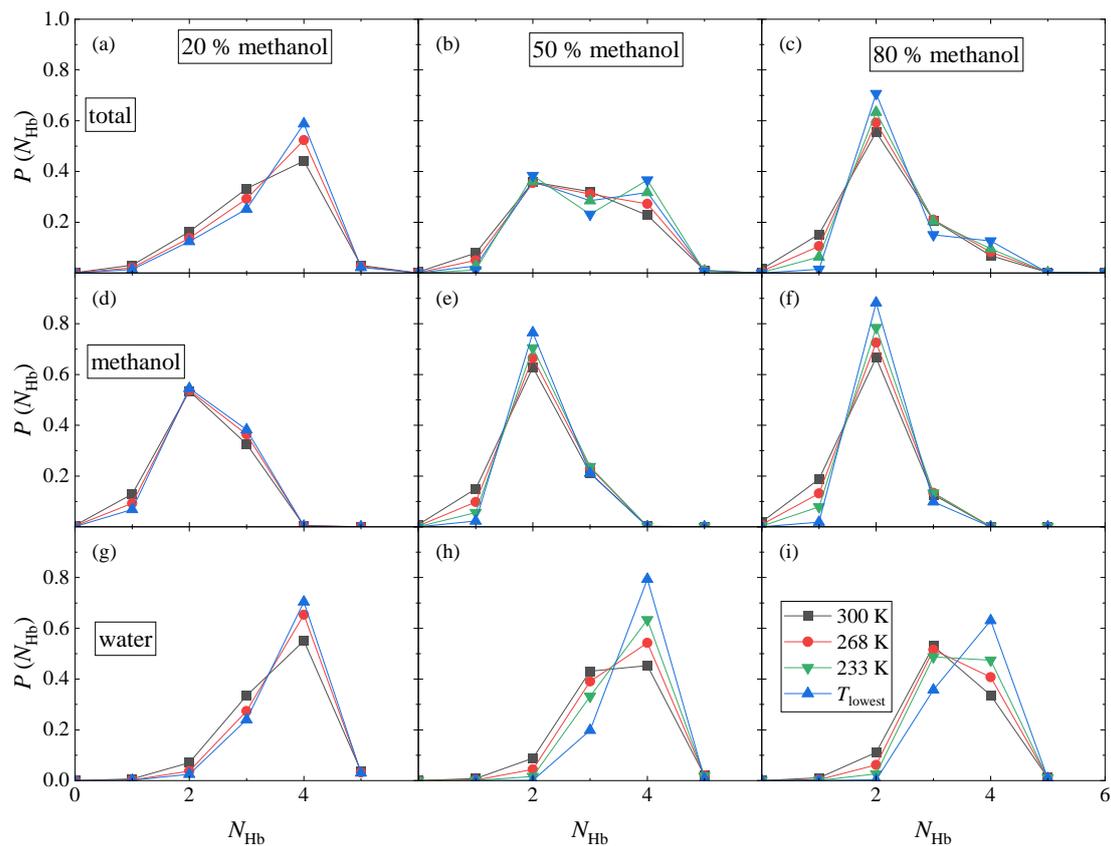

**Figure S5.** Distribution of the number of H-bonds at different temperatures and methanol concentrations in methanol-water mixtures concerning (top panels) any types of molecules, (middle panels, d-f) H-bonds of methanol molecules, (lower panels g - i) H-bonds of water molecules. The $T_{lowest}$ temperatures (the lowest investigated temperature at a given methanol concentration) are as follows: (left panels, a, d, g) $x_M$ = 0.2, 243 K; (middle panels, b, e, h) $x_M$ = 0.5, 193 K; (right panels, c, f, i) $x_M$ = 0.8, 163 K. (The lines are just guides to the eye.)

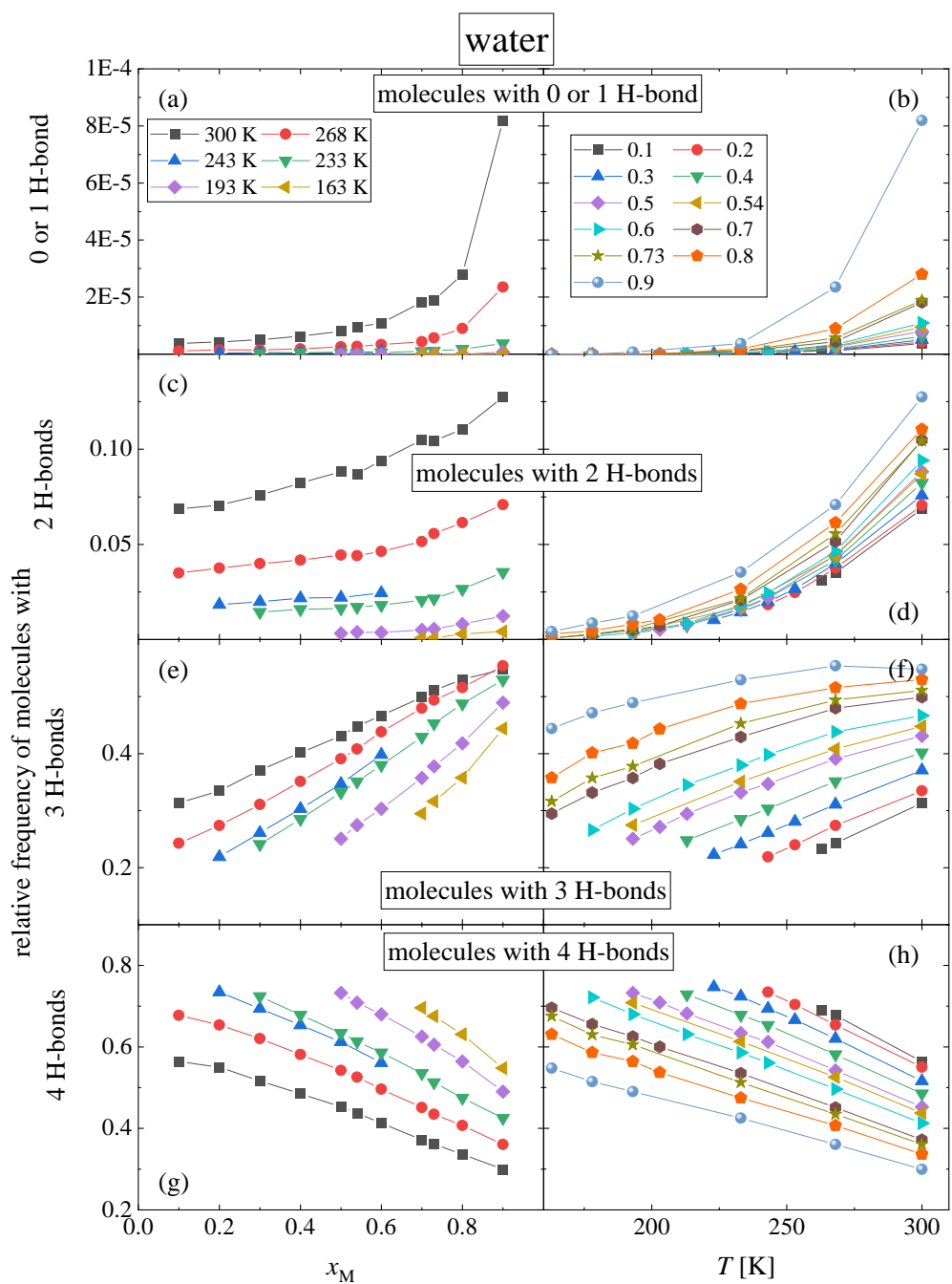

**Figure S6.** (left side) Concentration and (right side) temperature dependence of the relative frequencies of water molecules with (a, b) 0 or 1, (c, d) 2, (e, f) 3, and (g, h) 4 H-bonds. All types of bonds (both water – water and water – methanol bonds) are taken into account.

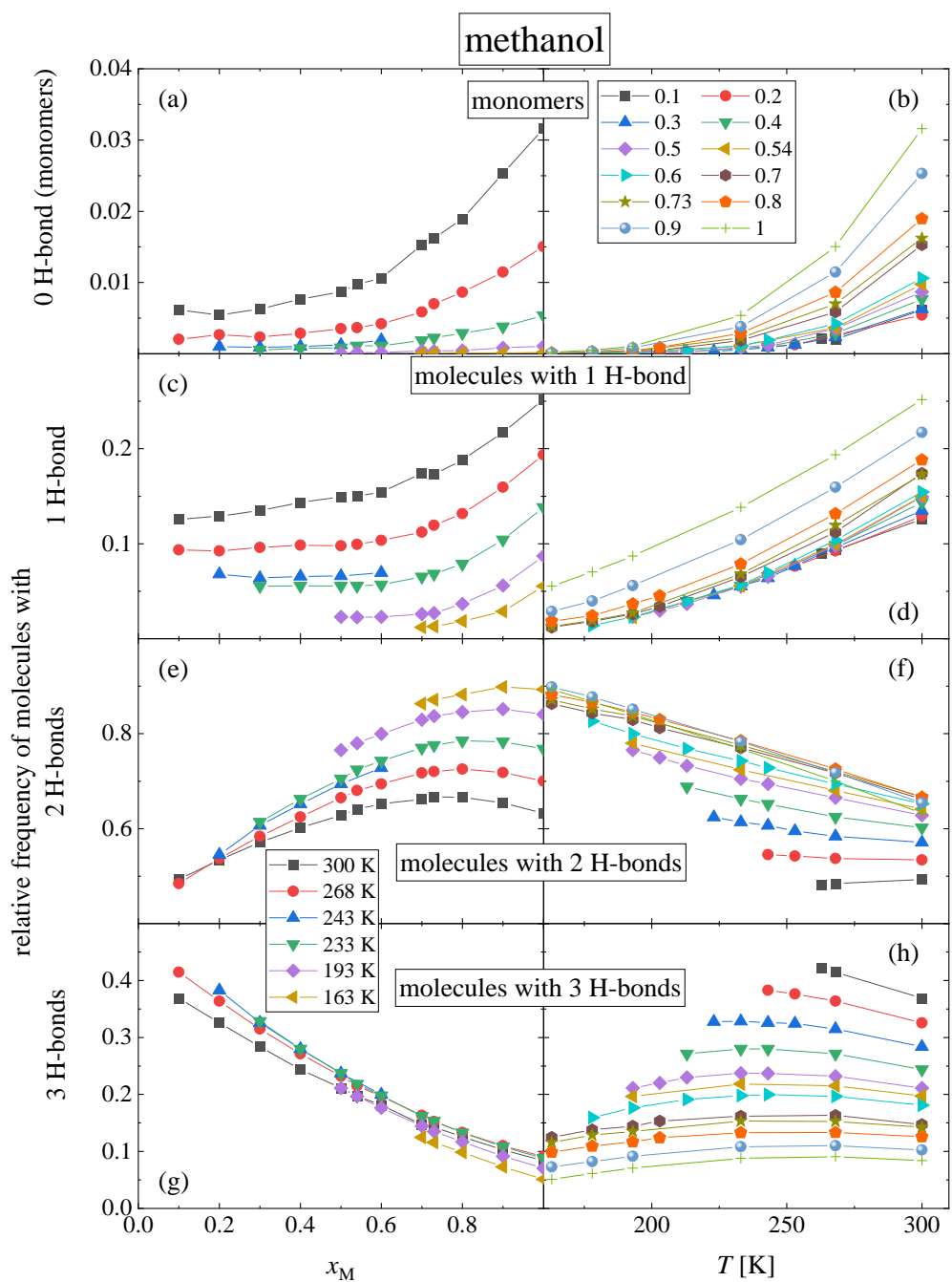

**Figure S7.** (a, c, e, g) Concentration and (b, d, f, h) temperature dependence of the relative frequencies of methanol molecules with (a, b) 0, (c, d) 1, (e, f) 2, and (g, h) 3 H-bonds. All types of bonds (both methanol – methanol and methanol – water bonds) are considered.

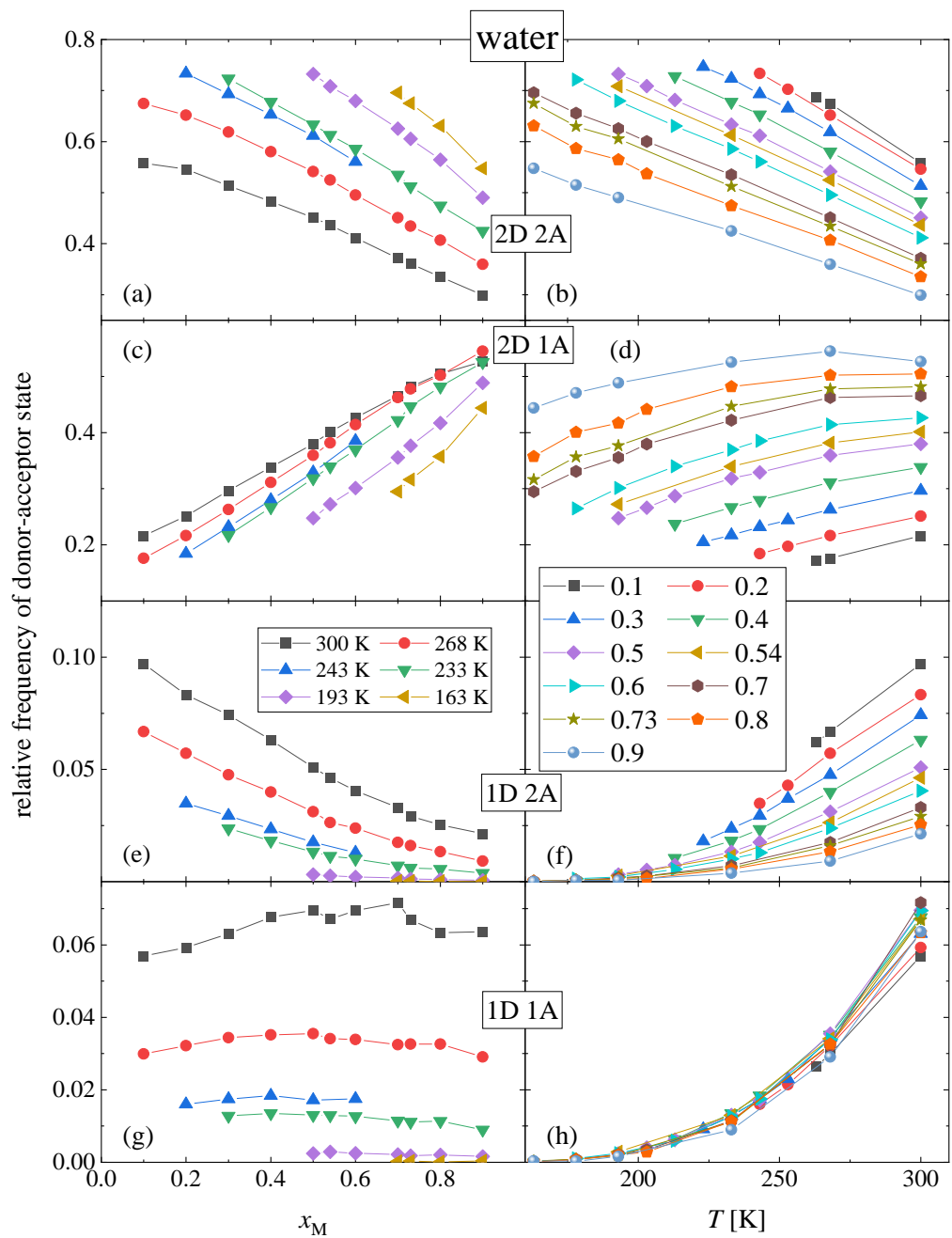

**Figure S8.** (left side) Concentration and (right side) temperature dependence of the most frequent donor-acceptor states of water molecules: (a, b) 2 donor, 2 acceptor state, (c, d) 2 donor, 1 acceptor state, (e, f) 1 donor, 2 acceptor state, and (g, h) 1 donor, 1 acceptor state.

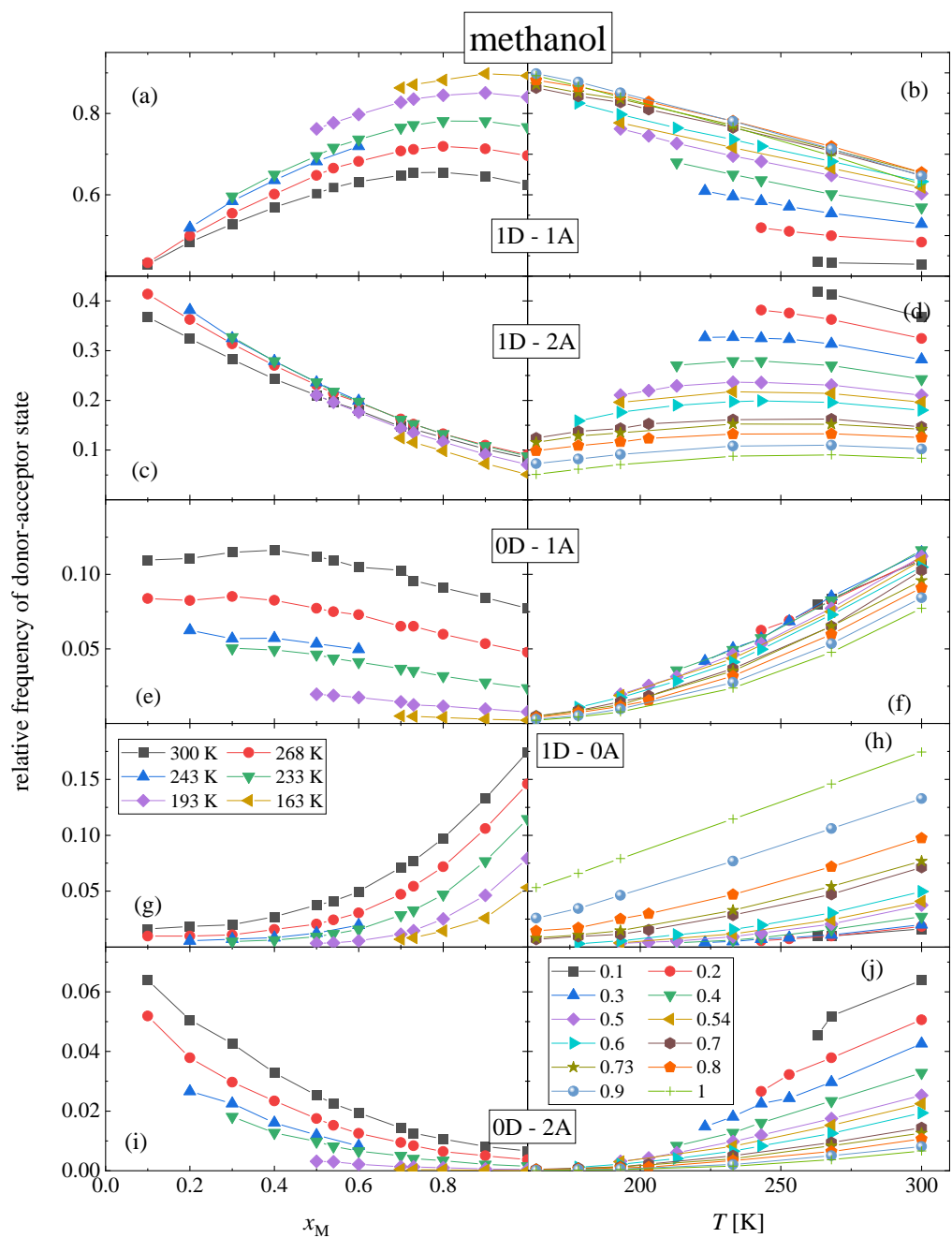

**Figure S9.** (left side) Concentration and (right side) temperature dependence of the most frequent donor-acceptor states of methanol molecules: (a, b) 1 donor, 1 acceptor state, (c, d) 1 donor, 2 acceptor state, (e, f) 0 donor 1, acceptor state, (g, h) 1 donor, 0 acceptor state, and (i, j) 0 donor, 2 acceptor state.

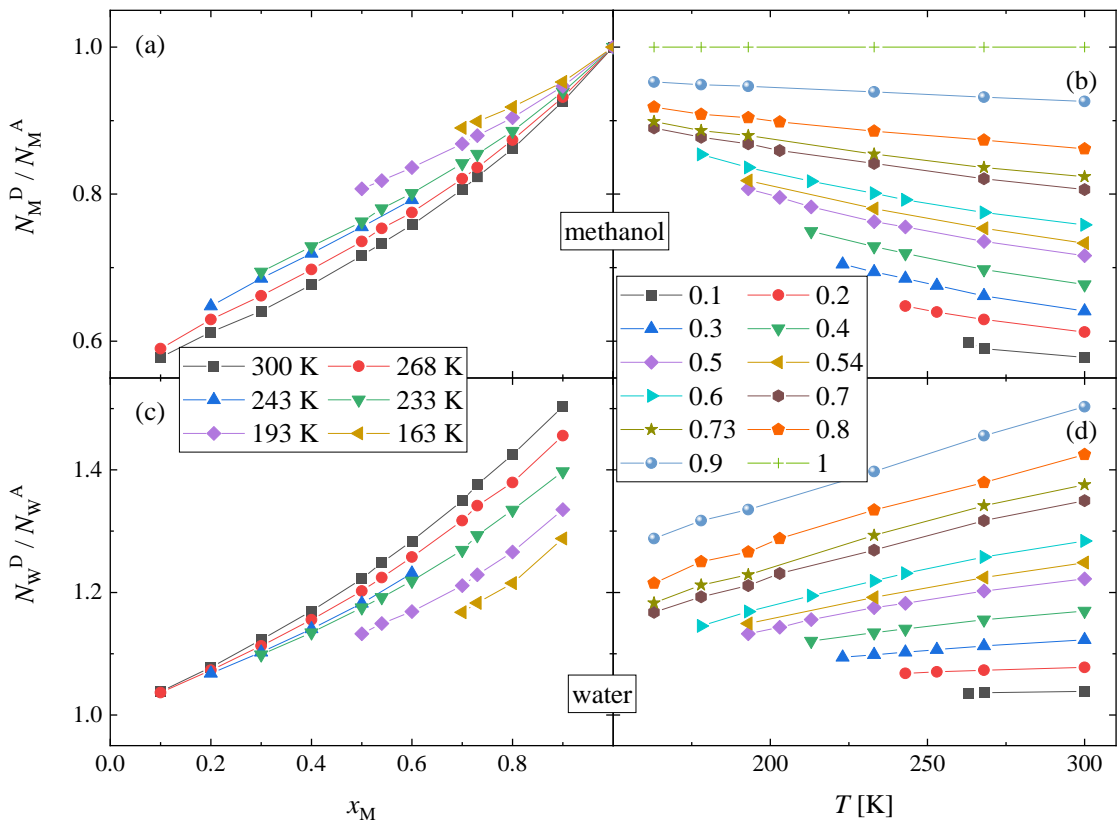

**Figure S10.** (left side) Concentration and (right side) temperature dependence of the donor/acceptor ratio of (a, b) methanol and (c, d) water molecules.

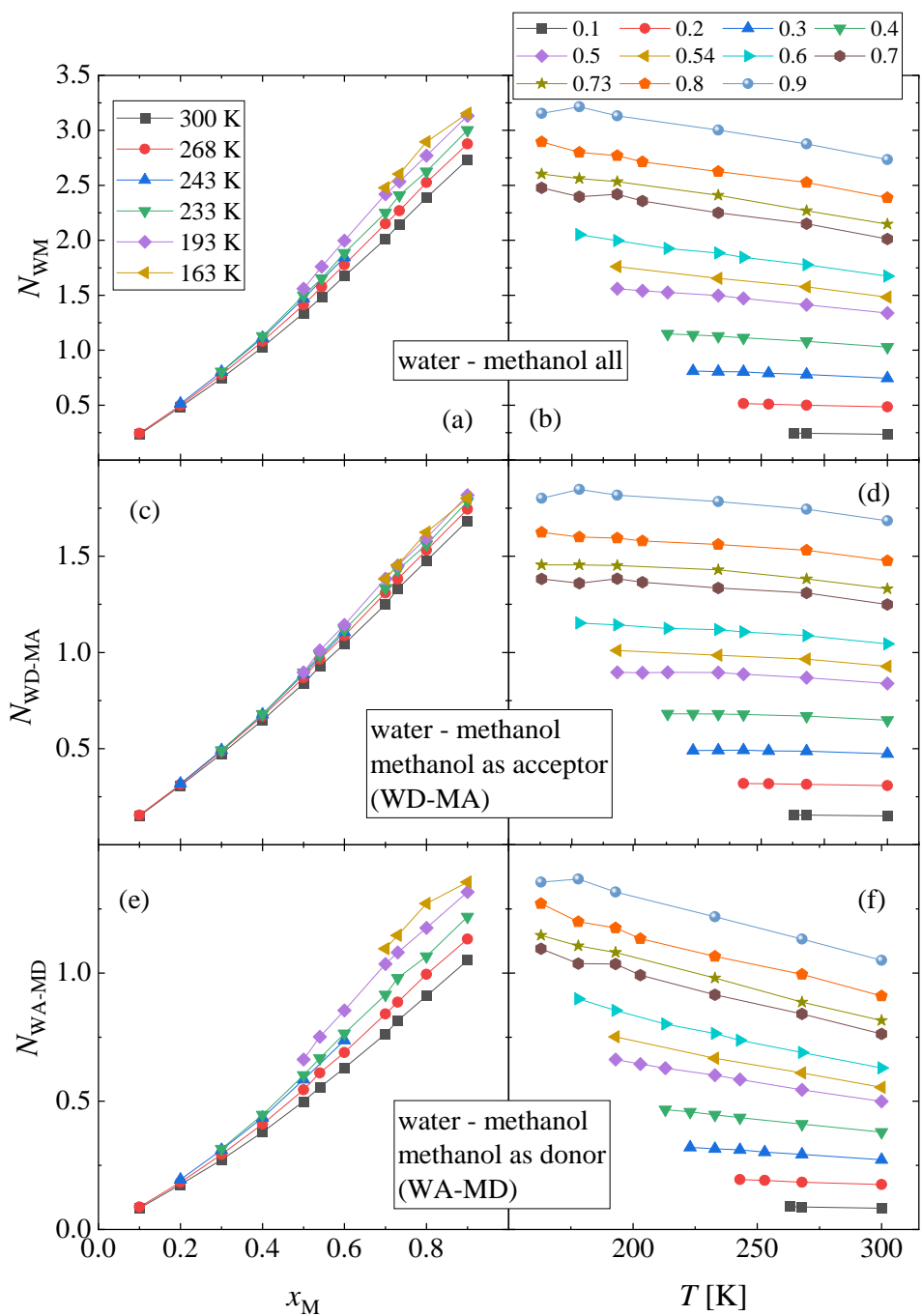

**Figure S11.** (Left panels) concentration and (right panels) temperature dependence of (a and b) the average number of H-bonded (donor and acceptor) methanol molecules around water, (c and d) the average number of H-bonded acceptor methanol molecules around water, (e and f) the average number of H-bonded donor methanol molecules around water.

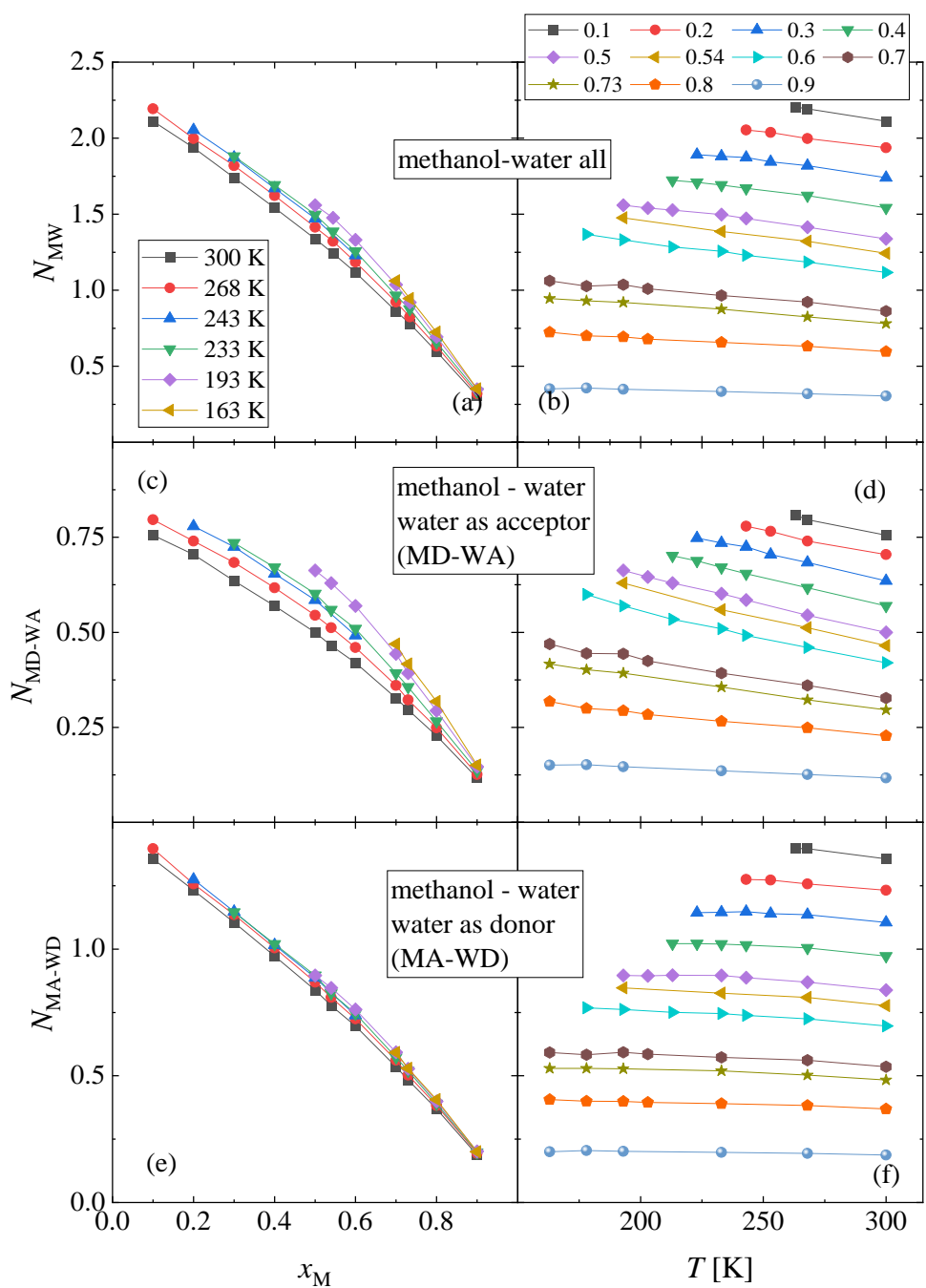

**Figure S12.** (Left panels) concentration and (right panels) temperature dependence of (a and b) the average number of H-bonded (donor and acceptor) water molecules around methanol, (c and d) the average number of H-bonded acceptor water molecules around methanol, (e and f) the average number of H-bonded donor water molecules around methanol.

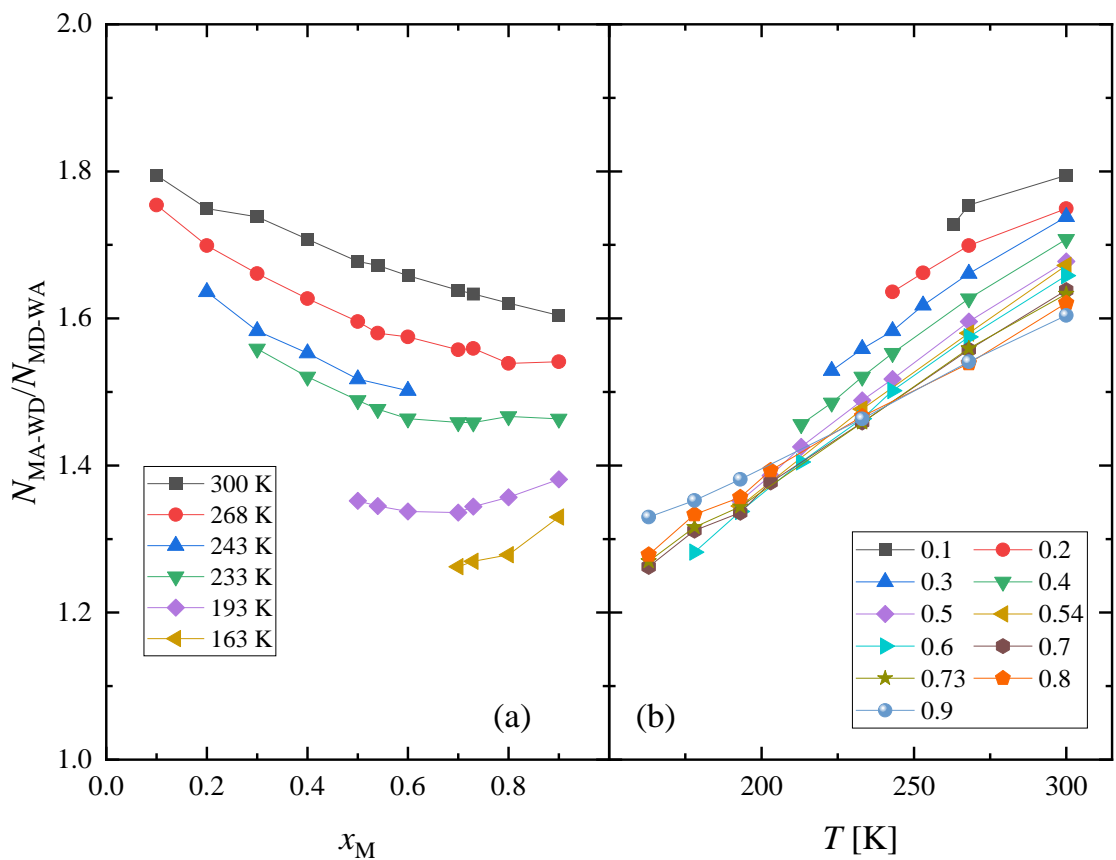

**Figure S13.** (a) Concentration and (b) temperature dependence of the ratio of methanol A – water D and methanol D – water A H-bonded pairs.

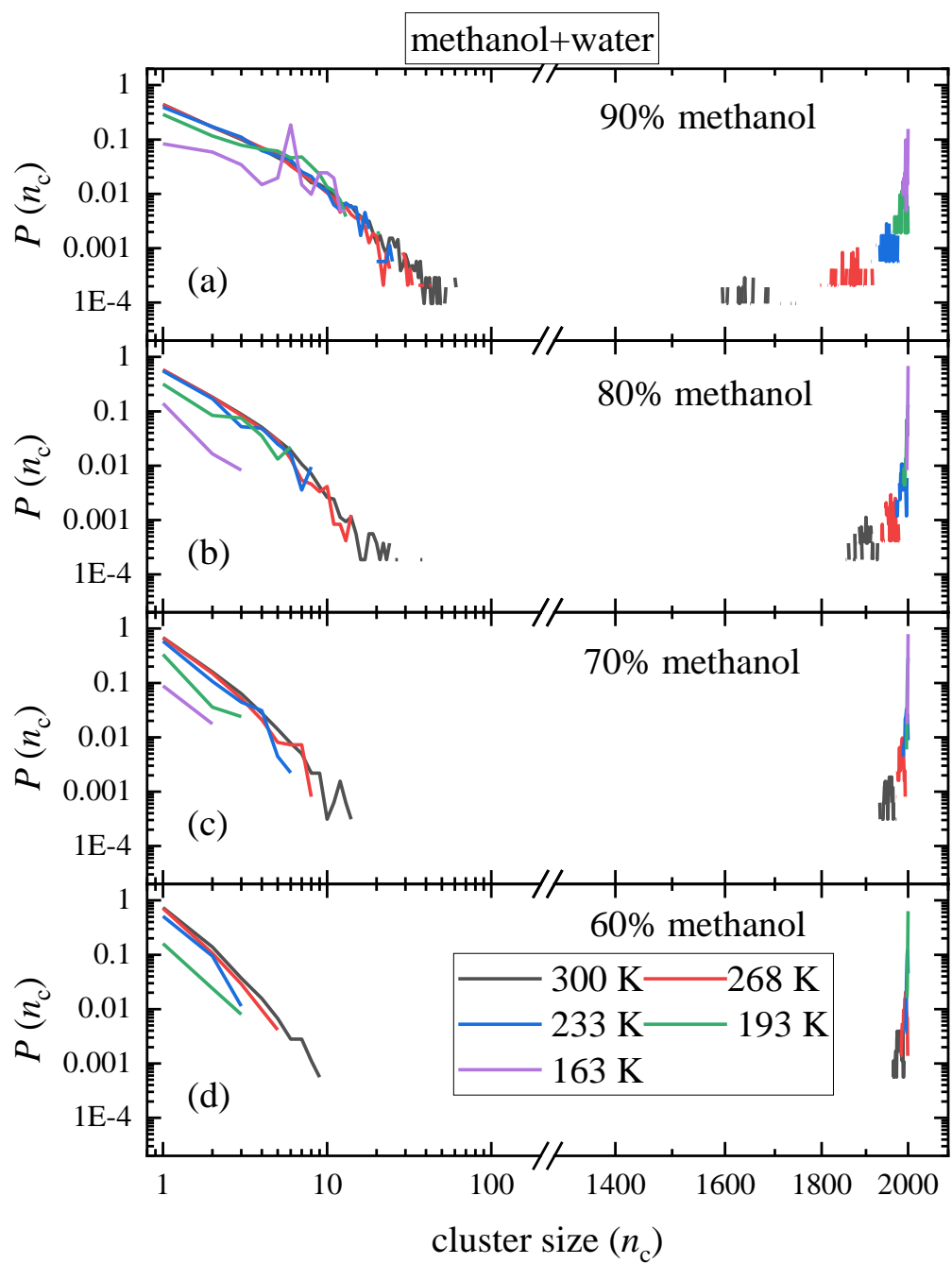

**Figure S14.** Cluster size distributions in alcohol-rich methanol – water mixtures at different temperatures, considering H-bonds between any types of molecules.

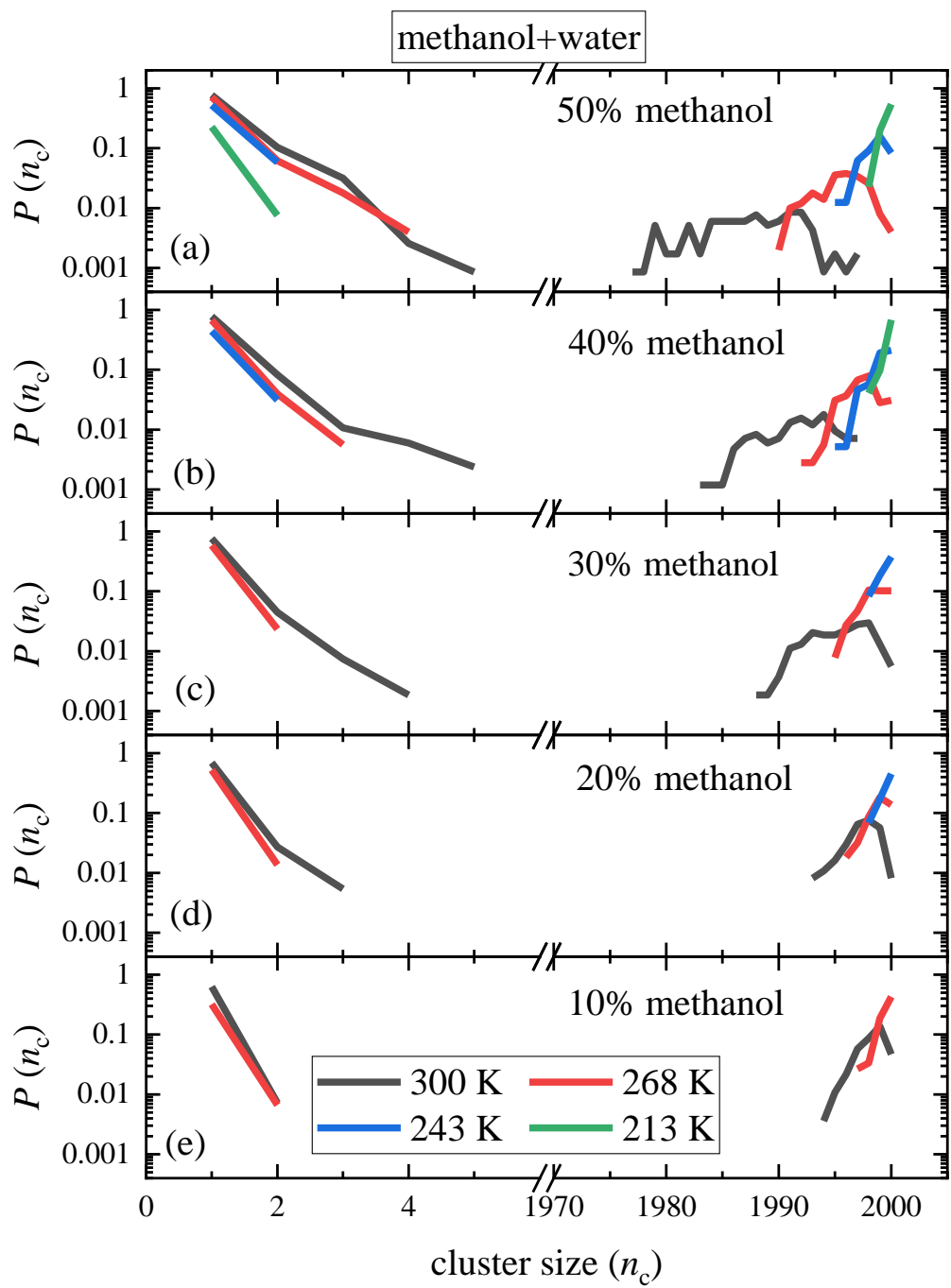

**Figure S15.** Cluster size distributions in water-rich and equimolar methanol – water mixtures at different temperatures, considering H-bonds between any types of molecules.

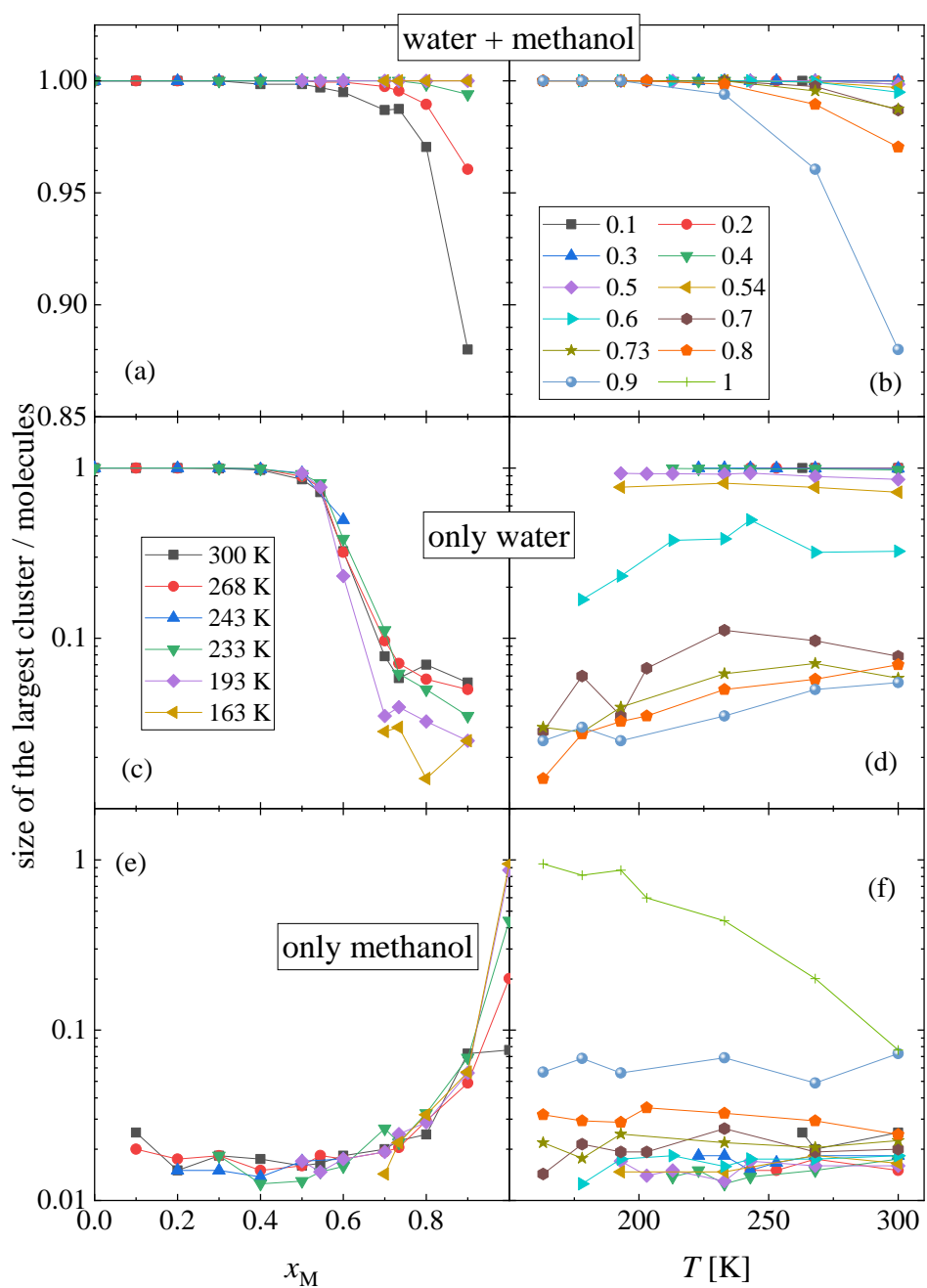

**Figure S16.** (left side, a, c, e) Concentration and (right side, b, d, f) temperature dependence of the size of the largest cluster considering (a, b) all molecules, (c, d) water subsystem, (e, f) methanol subsystem. Cluster sizes are normalized by the number of corresponding molecules in the simulation box. (Note that the *y*-axes of parts (a, b) are linear, while for the other parts, they are logarithmic.)

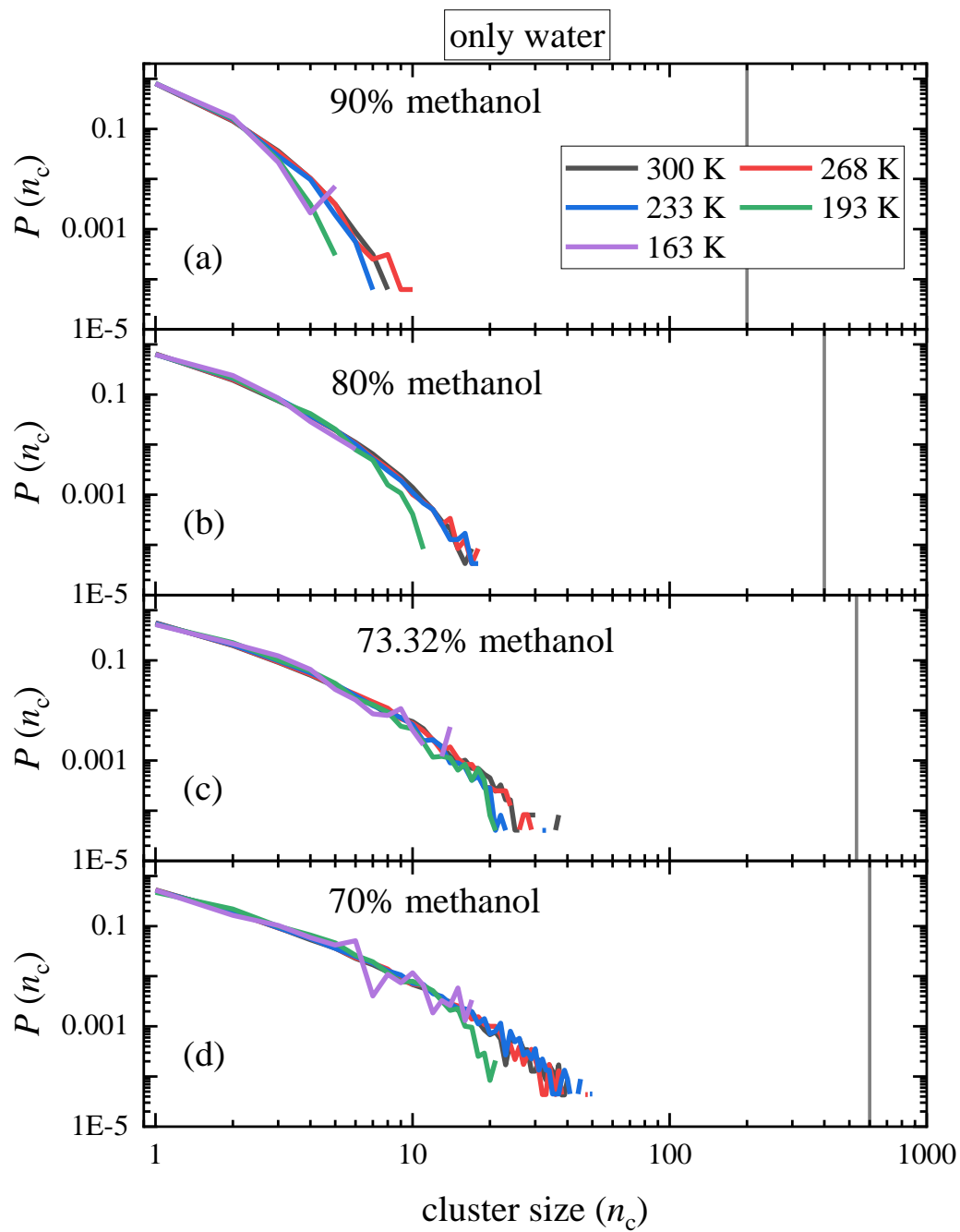

**Figure S17.** Cluster size distributions within the water subsystem in methanol-rich methanol – water mixtures at different temperatures (only H-bonds between water molecules are considered). Vertical lines show the numbers of water molecules in the simulation boxes.

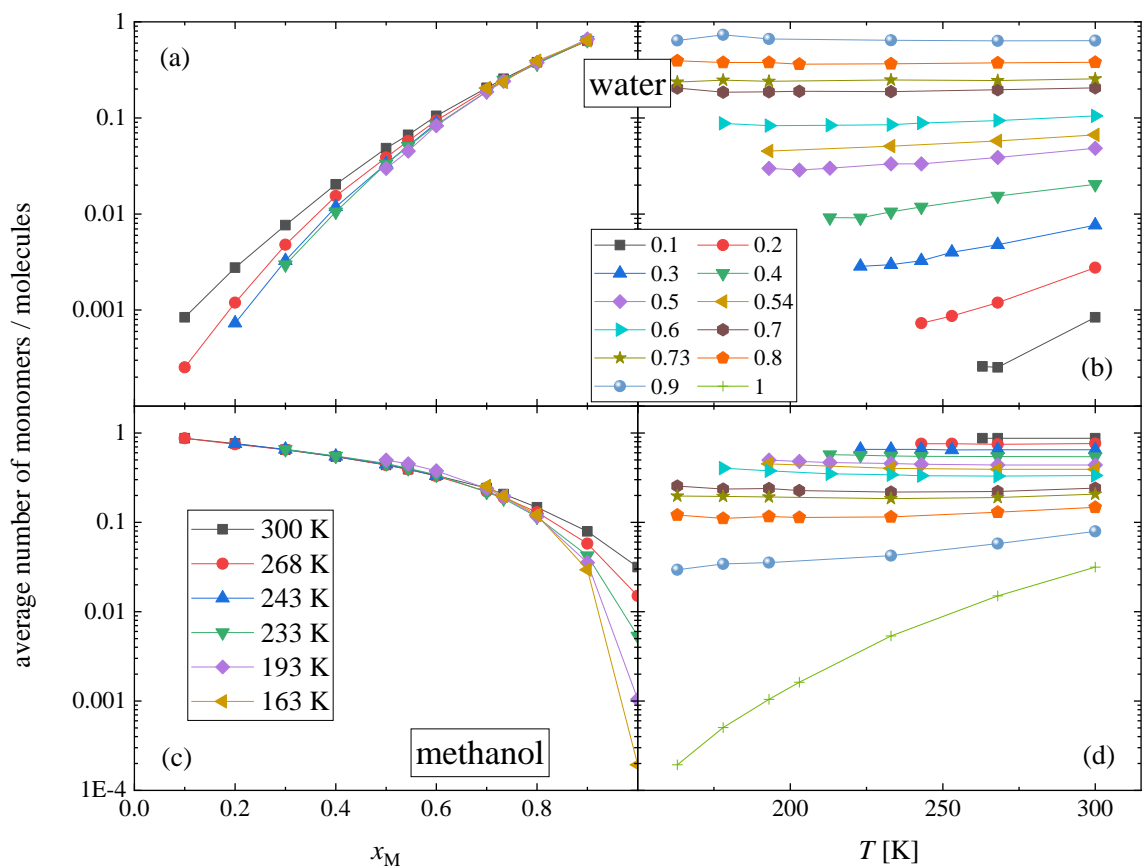

**Figure S18.** (left side, a, c) Concentration and (right side, b, d) temperature dependence of the average number of monomer molecules considering the (a, b) water subsystem, (c, d) methanol subsystem. The number of monomers is normalized by the number of corresponding molecules in the simulation box.

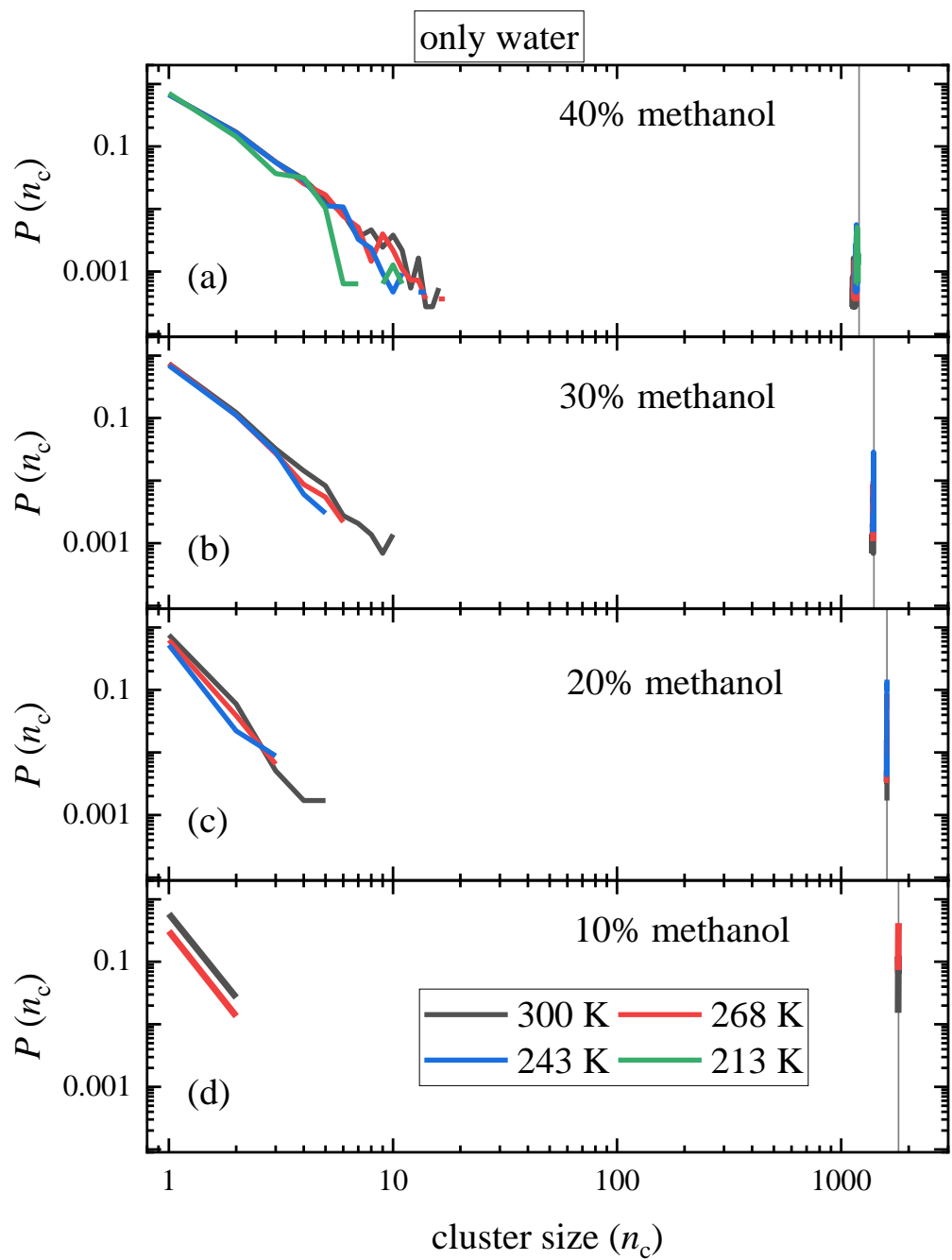

**Figure S19.** Cluster size distributions within the water subsystem in water-rich methanol – water mixtures at different temperatures (only H-bonds between water molecules are considered). Vertical lines show the numbers of water molecules in the simulation boxes.

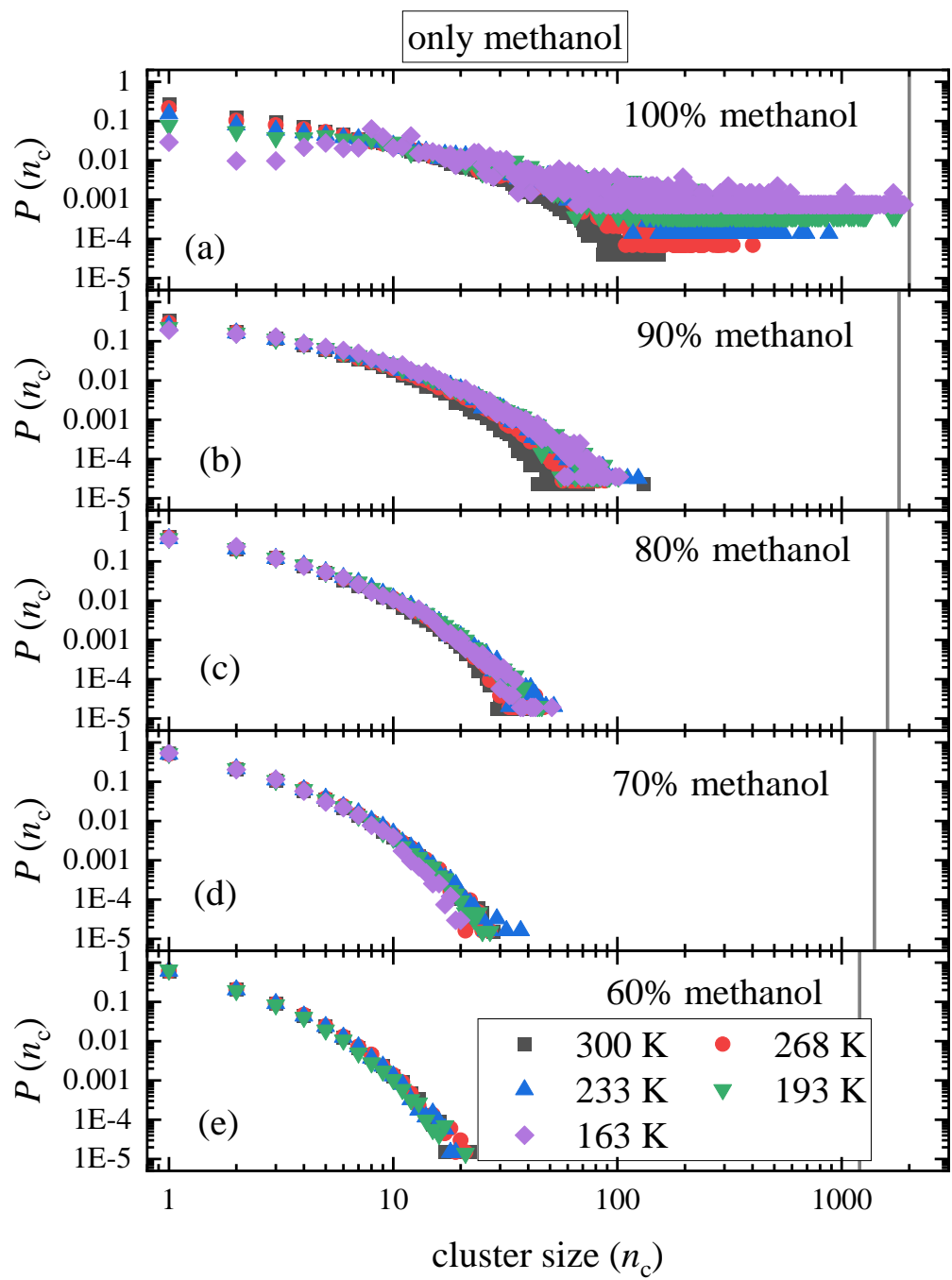

**Figure S20.** Cluster size distributions within the methanol subsystem in methanol-rich methanol – water mixtures at different temperatures (only H-bonds between methanol molecules are considered.) Vertical lines show the numbers of methanol molecules in the simulation boxes.

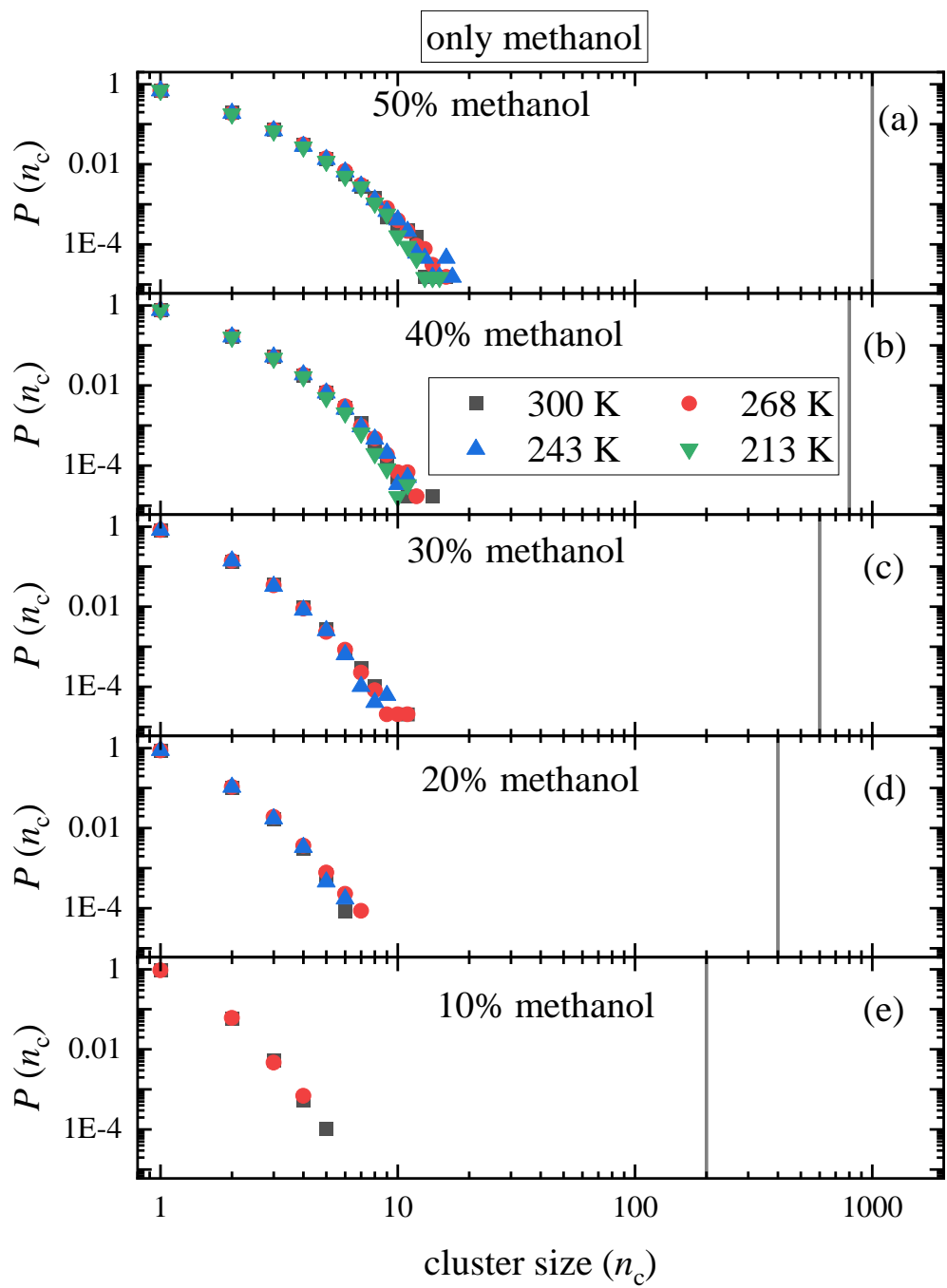

**Figure S21.** Cluster size distributions within the methanol subsystem in water-rich methanol – water mixtures at different temperatures (only H-bonds between methanol molecules are considered.) Vertical lines show the numbers of methanol molecules in the simulation boxes.

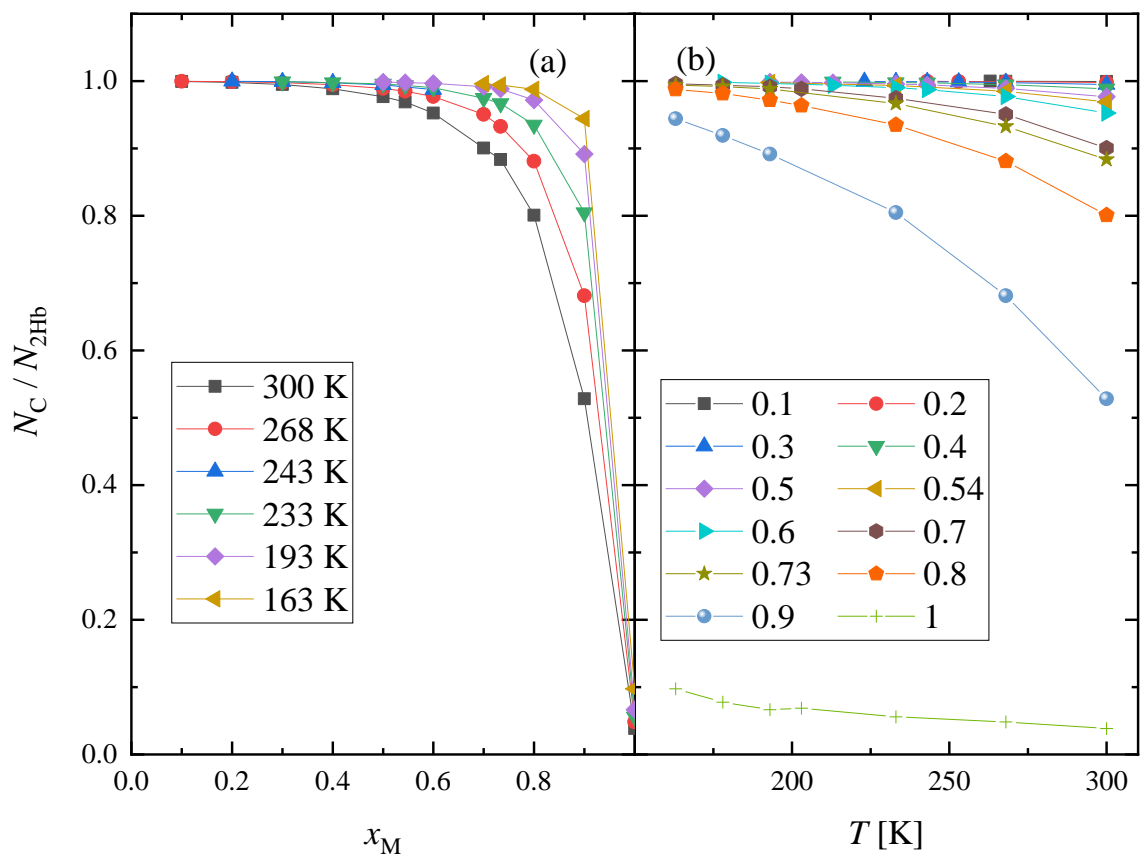

**Figure S22.** (a) Concentration and (b) temperature dependence of the ratio of molecules participating in cycles and molecules with at least 2 H-bonds. (All types of bonds and all molecules are taken into account.)

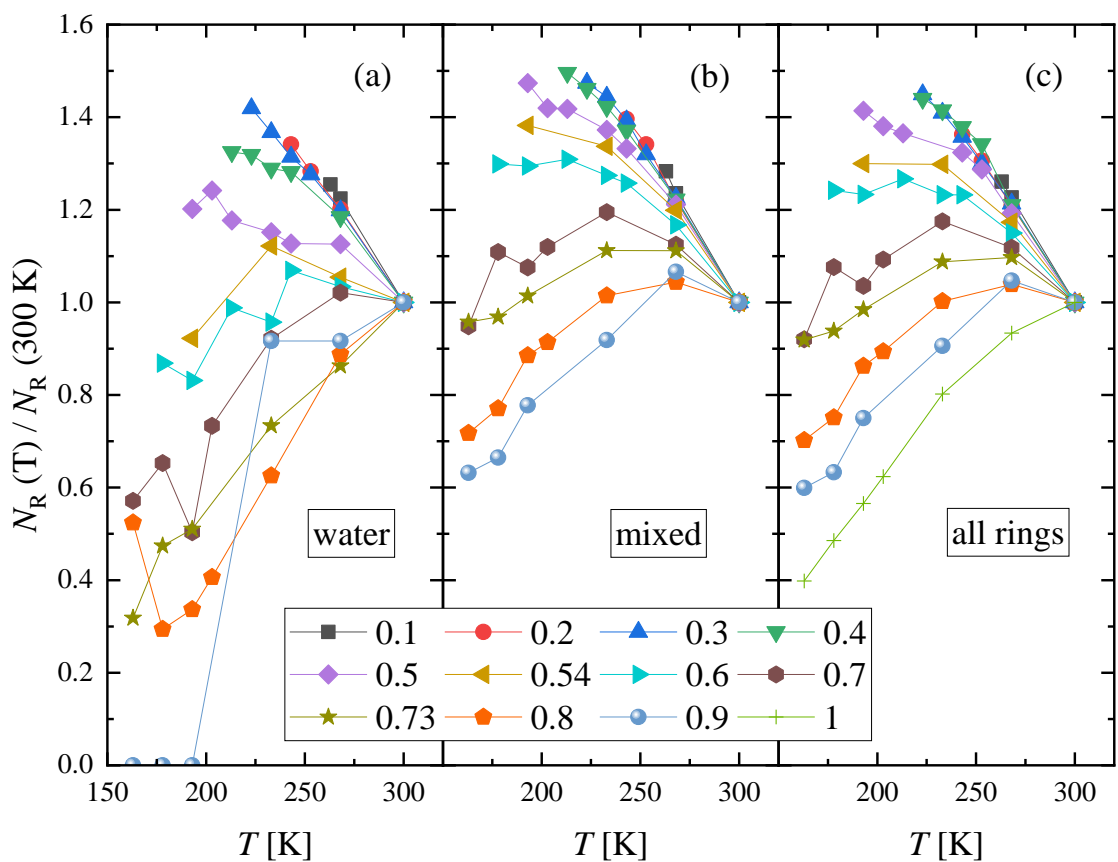

**Figure S23.** Temperature dependence of the number of (a) water, (b) mixed, (c) all (water, methanol, mixed) type rings normalized by the 300 K value. (Note: the number of the methanol rings is extremely low, not shown.)

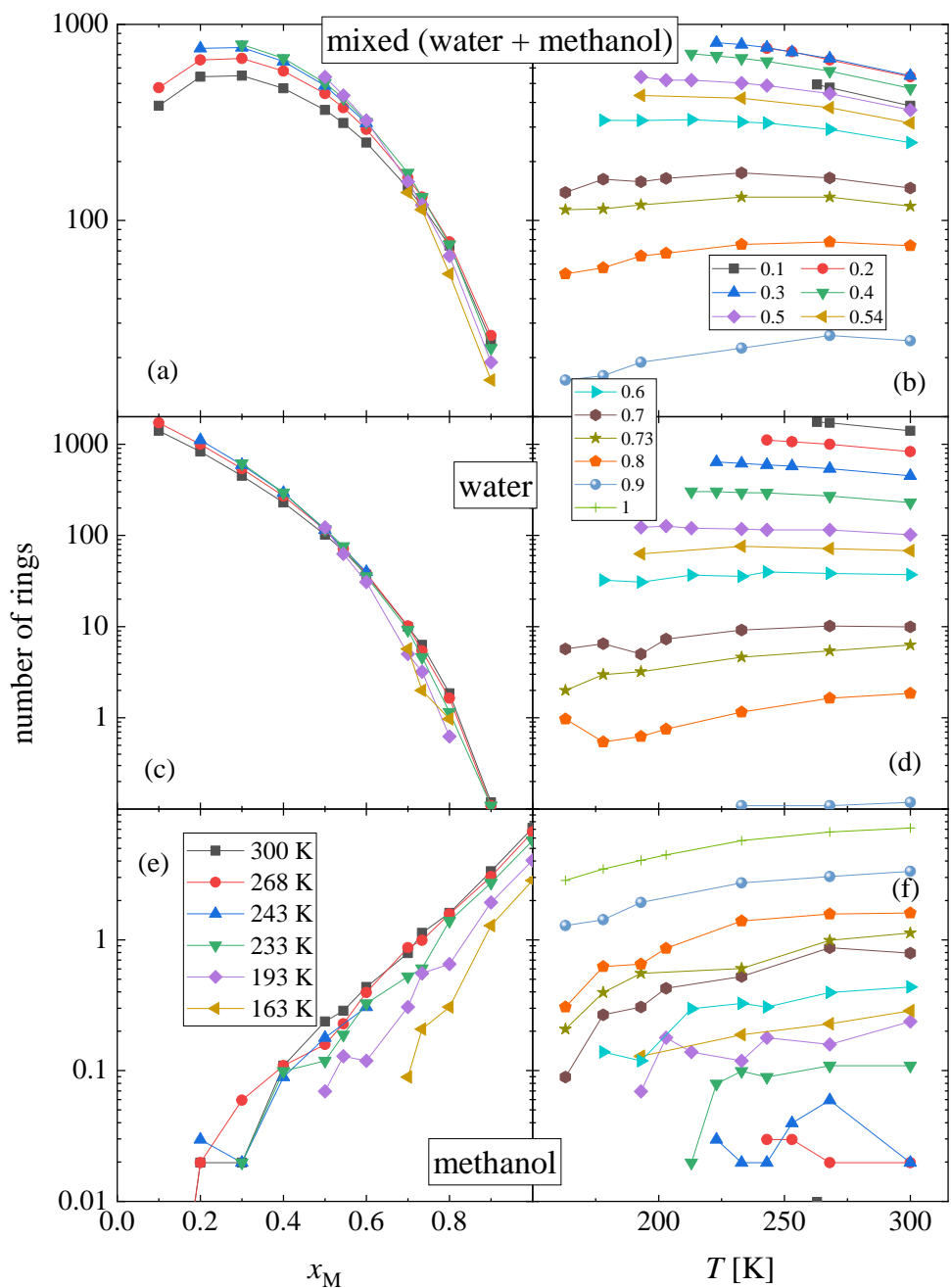

**Figure S24.** (Left panel) Concentration and (right panel) temperature dependence of the frequency of different ring types in methanol – water mixtures. Rings contain (a, b) both water and methanol (mixed rings), (c, d) only water, and (e, f) only methanol molecules.

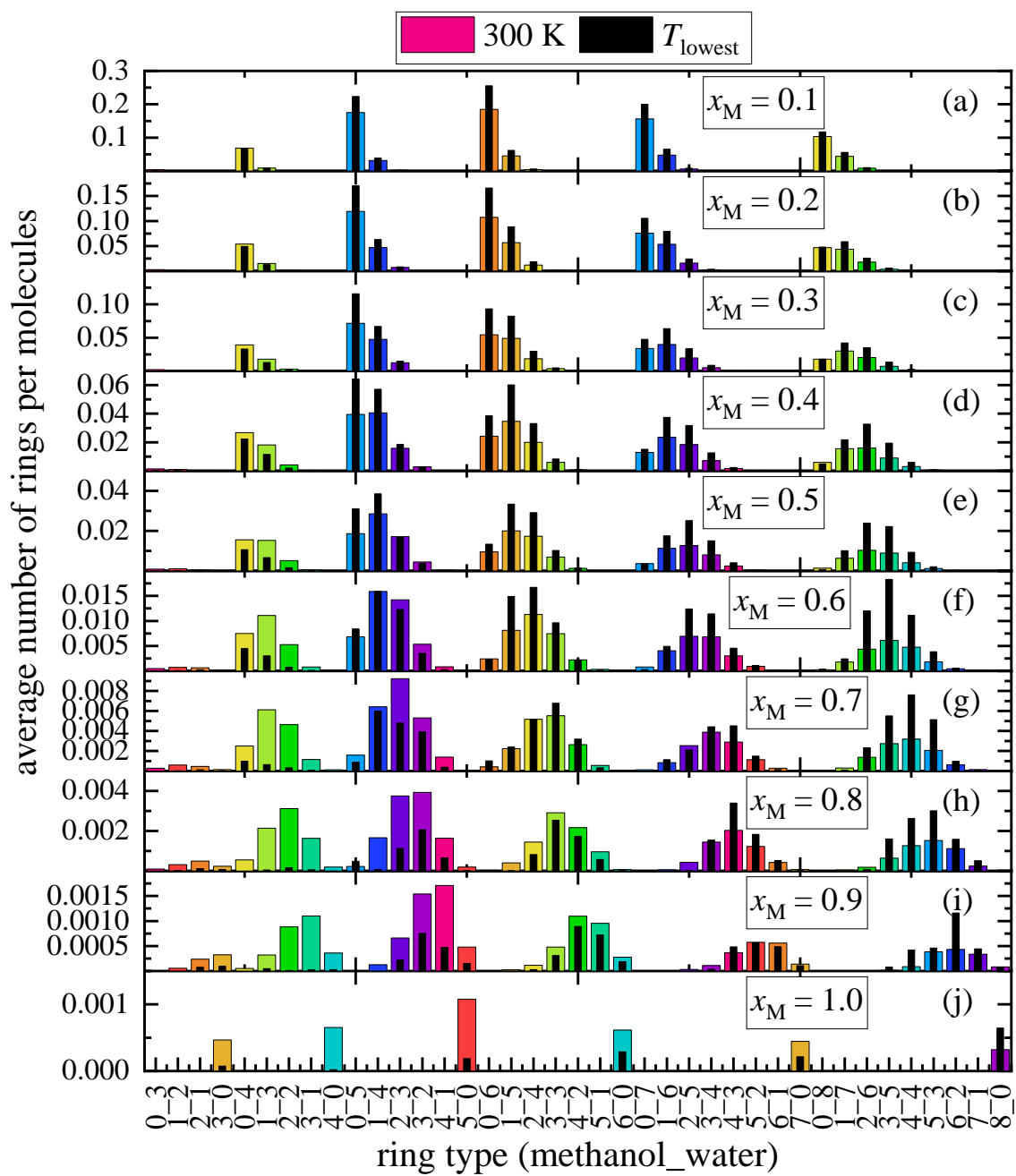

**Figure S25.** Temperature and concentration dependence of the ring type distributions in methanol – water mixtures. The number of rings is normalized by the number of molecules in the simulation box. The colored bars show the values at 300 K, whilst the black bars are the values at the lowest temperature investigated for the given composition: (a) $x_M$ = 0.1, 263 K; (b) $x_M$ = 0.2, 243 K; (c) $x_M$ = 0.3, 223 K; (d) $x_M$ = 0.4, 213 K; (e) $x_M$ = 0.5, 193 K; (f) $x_M$ = 0.6, 178 K; (g) – (j) 0.7 ≤ $x_M$ ≤ 1.0, 163 K.